\documentclass[reprint,prl,aps, floatfix]{revtex4-1}
\usepackage{graphicx}
\usepackage{color}
\usepackage{dcolumn}
\usepackage{bm}
\usepackage{amssymb}
\usepackage{braket}
\usepackage{color,amsmath}
\usepackage{comment}
\usepackage{gensymb}
\usepackage{soul,xcolor} 
\setstcolor{red} 
\usepackage{epstopdf}
\usepackage{hhline}
\usepackage{tabularx}
\usepackage{xspace}
\usepackage{float}
\usepackage{lineno} 

\newcolumntype{C}[1]{>{\centering\arraybackslash}m{#1}}
\newcolumntype{N}{@{}m{0pt}@{}}

\definecolor{cadmiumgreen}{rgb}{0.0, 0.42, 0.24}

\newcommand{\moire}{moir\'e\xspace}

\begin{document}

\title{On the origin of anomalous hysteresis in graphite/boron nitride transistors}

\author{Dacen Waters$^{1,2*}$}
\author{Derek Waleffe$^{1*}$}
\author{Ellis Thompson$^{1}$}
\author{Esmeralda Arreguin-Martinez$^{3}$}
\author{Jordan Fonseca$^1$}
\author{Thomas Poirier$^4$}
\author{James H. Edgar$^4$}
\author{Kenji Watanabe$^{5}$} 
\author{Takashi Taniguchi$^{6}$} 
\author{Xiaodong Xu$^{1,3}$}
\author{David Cobden$^{1\dagger}$}
\author{Matthew Yankowitz$^{1,3\dagger}$}

\affiliation{$^{1}$Department of Physics, University of Washington, Seattle, Washington, 98195, USA}
\affiliation{$^{2}$ Intelligence Community Postdoctoral Research Fellowship Program, University of
Washington, Seattle, Washington, 98195, USA}
\affiliation{$^{3}$Department of Materials Science and Engineering, University of Washington, Seattle, Washington, 98195, USA}
\affiliation{$^{4}$Tim Taylor Department of Chemical Engineering, Kansas State University, Durland Hall, Manhattan, KS, USA}
\affiliation{$^{5}$Research Center for Electronic and Optical Materials, National Institute for Materials Science, 1-1 Namiki, Tsukuba 305-0044, Japan}
\affiliation{$^{6}$Research Center for Materials Nanoarchitectonics, National Institute for Materials Science, 1-1 Namiki, Tsukuba 305-0044, Japan}
\affiliation{$^{*}$These authors contributed equally to this work.}
\affiliation{$^{\dagger}$ cobden@uw.edu (D.C.); myank@uw.edu (M.Y.)}

\maketitle

\textbf{Field-effect devices constructed by stacking flakes of van der Waals (vdW) materials, with hexagonal boron nitride (hBN) playing the role of gate dielectric, often exhibit virtually no hysteresis in their characteristics. This permits exquisitely detailed studies of diverse gate-voltage-tuned phenomena in vdW devices. Recently, however, a dramatic form of gate hysteresis, sometimes called the “gate doesn’t work” (GDW) or “electron ratchet” effect, has been seen in certain individual vdW devices that seem otherwise unexceptional~\cite{Zheng2020,Niu2022,Wang2022hBNintercalated,Klein2023,Ren2023,zheng2023electronicratcheteffectmoire,zhang2024MLG,chen2024tDBLG,niu2024tMoS2,Chen2024tBLGneuromorphiccomp}. When it occurs, this hysteresis phenomenon is striking and robust, yet it is difficult to reliably reproduce between devices and, largely as a result, its origin remains disputed. Most devices where it has been seen have a bilayer graphene channel and nominal rotational alignment between the graphene and hBN, which has engendered explanations based on properties of bilayer graphene combined with moir\'e effects~\cite{Zheng2020,Niu2022,zheng2023electronicratcheteffectmoire}. Here, we report our studies of the phenomenon observed in devices that have multilayer graphene channels. We find that the effect can occur in devices with graphite channels that have many more than two graphene layers, in which case it is unambiguously associated with just one surface of the graphite. It can also survive to room temperature, occur in the absence of intentional rotational alignment with hBN, persist when a monolayer of WSe$_2$ is inserted between the graphene and hBN, and exhibit continuous relaxation on timescales of hours or longer. These observations impose strong constraints on the origin of this puzzling phenomenon, which has exciting potential applications if it can be mastered~\cite{Yan2023,Klein2023,Chen2024tBLGneuromorphiccomp}.}

In a field-effect transistor (FET), the resistance of the channel is modulated by a voltage applied to a metallic gate separated from it by a dielectric layer. Hysteresis will occur in the channel resistance versus gate voltage if the changing gate electric field causes a rearrangement of charge in the dielectric. The mechanisms for this, which reflect various ways in which the dielectric is not “ideal”, include electrons moving between localized states in the bulk and interfaces of the dielectric~\cite{Gu2005_OTFThyst, Burke2012_GaAshyst}, ferroelectric domain reversal, ion drift, and other structural changes. FETs made from stacked two-dimensional (2D) van der Waals materials with hBN as the dielectric and graphite gate electrodes often show remarkably little, if any, gate hysteresis. Although hysteresis can be induced by optical illumination~\cite{Ju2014}, which promotes tunneling of electrons to or from defect states in the hBN (with similar effects also driven by STM voltage pulses or electron-beam irradiation~\cite{Velasco2016,Wong2015,Shi2020}), it seems that under the right conditions hBN can act as an effectively ideal dielectric.  

\begin{figure*}[t]
\includegraphics[width=\textwidth]{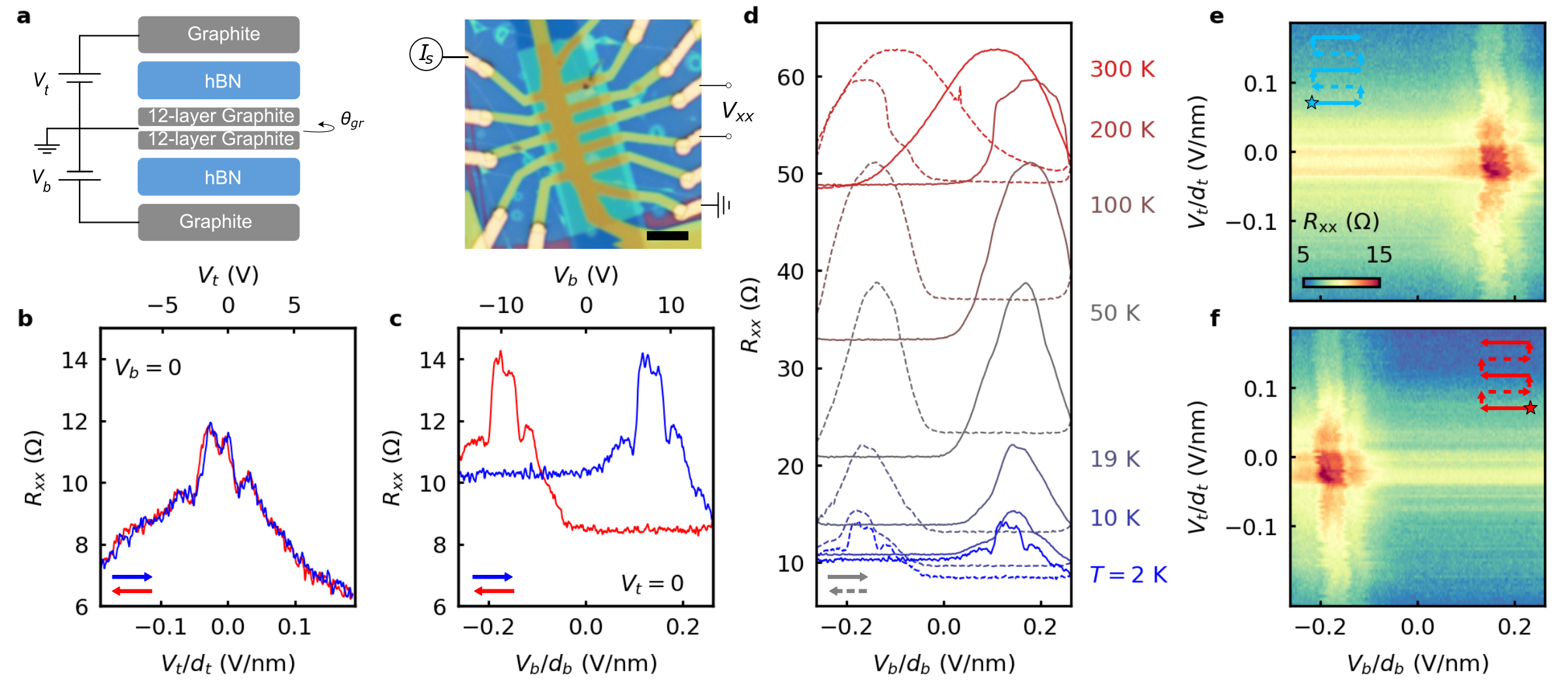} 
\caption{\textbf{Hysteresis in a 24-layer graphite device.}
\textbf{a}, (left) Cartoon schematic of a 24-layer Bernal graphite device, comprising two 12-layer graphite flakes stacked with a twist angle $\theta_{\rm gr}=1.21^{\circ}$. The top ($V_t$) and bottom ($V_b$) graphite gates have hBN dielectrics of thickness $d_t$ and $d_b$, respectively. 
(right) Optical micrograph of the device, with annotations showing the measurement setup. Here, $I_s$ is the a.c. source current and $V_{xx}$ is the measured potential difference, such that $R_{xx}=V_{xx}/I_s$. Scale bar is $5~\rm \mu m$.
\textbf{b-c}, Longitudinal resistance, $R_{xx}$, as a function of top (\textbf{b}) and bottom (\textbf{c}) gate bias. The top axes show $V_t$ and $V_b$. On the bottom axes, the gate voltages are normalized by the thickness of their respective hBN dielectrics. The blue and red arrows indicate the gate sweep direction. In both measurements, the other gate is held at ground.
\textbf{d}, $R_{xx}$ as a function of $V_b$ at several selected temperatures. Each curve is taken with $V_t=0$. 
\textbf{e-f}, Maps of $R_{xx}$ as a function of both $V_t$ and $V_b$, where $V_b$ is swept from negative to positive (\textbf{e}) or from positive to negative (\textbf{f}). The maps are taken by sweeping $V_b$ as the fast axis and $V_t$ as the slow axis (negative to positive). The arrows show the relevant gate sweeping directions, where data is collected only for the sweeps corresponding to solid arrows. The stars indicate the starting point of the measurement.
}
\label{fig:1}
\end{figure*}

Recently, however, a dramatic and peculiar kind of gate hysteresis has been reported in a number of individual vdW FET devices. In many of these devices, the channel is Bernal-stacked bilayer graphene in which one of the encapsulating hBN flakes is in close rotational alignment with the graphene~\cite{Zheng2020,Niu2022,zheng2023electronicratcheteffectmoire,Yan2023}. However, similar behavior has also been reported in devices when the channel is magic-angle twisted bilayer graphene~\cite{Klein2023,Chen2024tBLGneuromorphiccomp}, twisted double bilayer graphene~\cite{chen2024tDBLG}, and even in devices consisting of two graphene layers separated by monolayer hBN~\cite{Wang2022hBNintercalated}. Common factors to all reported instances of the phenomenon are that it is mostly associated with a single gate; that it is approximately symmetric about zero gate voltage; and that it is typically an “all or nothing” effect, in that it is either large or completely absent (though there have been instances of weak hysteresis of a possibly different nature). The most puzzling common feature, however, is the combination of behaviors referred to as the ``gate doesn’t work'' (GDW) and “electron ratchet” effects~\cite{Zheng2020,zheng2023electronicratcheteffectmoire}. As the gate voltage is swept the resistance is nearly constant over some range (i.e., the GDW effect), but eventually starts to change and sweeps out the anticipated transfer curve of the device. If the sweep direction is then reversed, no matter at what point, the resistance stops changing once again (i.e., the electron ratchet effect). This peculiar behavior has prompted theoretical explanations ranging from exotic correlated electron effects to interlayer sliding ferroelectric mechanisms~\cite{Zheng2020,Chen2024tBLGneuromorphiccomp,niu2024tMoS2,zheng2023electronicratcheteffectmoire, Niu2022,chen2024tDBLG}. 

Here, we report the observation of the GDW/ratchet effect in several new device geometries. Taken together, they shed light on the possible origin of the hysteresis effects. Specifically, we see the hysteresis in devices with much thicker graphite channels (up to 24 layers), showing that having just a few layers of graphene as the channel is not necessary, and also that the hysteresis can be unambiguously connected with the electric field on just one side of the channel. The hysteresis is almost unchanged up to room temperature, ruling out mechanisms with low energy scales. We see it in devices with no known rotational alignment to the hBN on the side with the hysteretic gate, indicating that moir\'e patterns are also not essential. Further, we see it in a device where monolayer WSe$_2$ is inserted between the graphene and hBN, implying that direct contact between graphene and hBN is also not crucial. In this device, we establish a gradual partial relaxation of the resistance on a time scale of hours, implying that there exists a continuum of history-dependent configurations. After presenting these findings we give a detailed discussion arguing that we can exclude strongly correlated effects and sliding ferroelectricity as the principle underlying mechanisms, under the plausible assumption of a common origin for the phenomenon across all studies. We then consider a scenario involving charges moving within the hBN which qualitatively explains all facets of this peculiar phenomenon, with the important proviso that it requires a distribution of charge traps in the hBN that we cannot yet explain or intentionally reproduce. Overall, our results raise sharp questions about the dynamics of electrons in vdW dielectrics that deserve attention for their importance in understanding the GDW/ratchet effect and its employment in applications.
  
\medskip\noindent\textbf{Hysteresis in a device with a graphite channel}

We first examine a device with a channel of 24-layer thick graphite, comprising two 12-layer Bernal-stacked flakes stacked on each other with twist angle of $1.21^{\circ}$ (Fig. 1a). As usual, the channel is encapsulated above and below with hBN, and the entire structure is surrounded by graphite top and bottom gates. Figures 1b-c show measurements of the four-terminal resistance, $R_{xx}$, at a temperature of 1.7 K as a function of the voltage on the top gate, $V_t$, and bottom gate, $V_b$. Electric fields being key, we divide each gate voltage by the thickness of the respective hBN. On sweeping $V_t$ up and down at fixed $V_b$ (Fig. 1b), we see a resistance peak feature near $V_t=0$ with negligible hysteresis. There is a similar resistance peak for the same measurement made with $V_b$, but in this case there is large hysteresis and the peak is offset from $V_b=0$ by an amount $\Delta V_b \approx$ 8.5~V. The offset is roughly equal and opposite in the two sweep directions, and nearly independent of the sweep speed. The value of $\Delta V_b/d_b\approx0.15\ \rm V/nm$ is a large fraction of the typical breakdown field of hBN, and the full width of the hysteresis region corresponds to a nominal capacitive change in the carrier density of the bottom accumulation layer of $\approx 7 \times 10^{12}$~cm$^{-2}$ (see Methods). The GDW effect is exemplified by the regions in Fig.~\ref{fig:1}c where sweeping $V_b$ in one direction appears to have no effect on the resistance, implying that no free carriers are accumulating in the channel. Figure 1d shows that the size of $\Delta V_b$ barely changes at temperatures up to 200~K, and even at 300~K it is still large although the behavior begins to look more like conventional hysteresis, i.e., with a much less pronounced region of the GDW effect. This phenomenon appeared unexpectedly in this particular device, as we did nothing in the fabrication process to produce it intentionally.

The magnitude of the hysteresis is independent of the value of $V_t$, as can be seen in Fig. 1e-f which shows maps of $R_{xx}$ made by sweeping $V_b$ either up (Fig. 1e) or down (Fig. 1f) and stepping $V_t$ as the slow axis. The resistance of such a thick graphite flake is sensitive to gate voltage because the accumulation of screening electrons or holes in the outermost graphene layers at its surface produces a sheet of lower resistivity that shunts the total resistance~\cite{Waters_Nature_2023}. A resistance peak is seen as either of the surfaces passes through its neutral state. The cross-like features in Fig. 1e-f occur because the charge accumulation on the bottom surface depends only on the voltage on the bottom gate and is fully screened from the effect of the top gate by the bulk of the graphite, and vice versa for the top surface (see Methods).

\begin{figure*}[t]
\includegraphics[width=\textwidth]{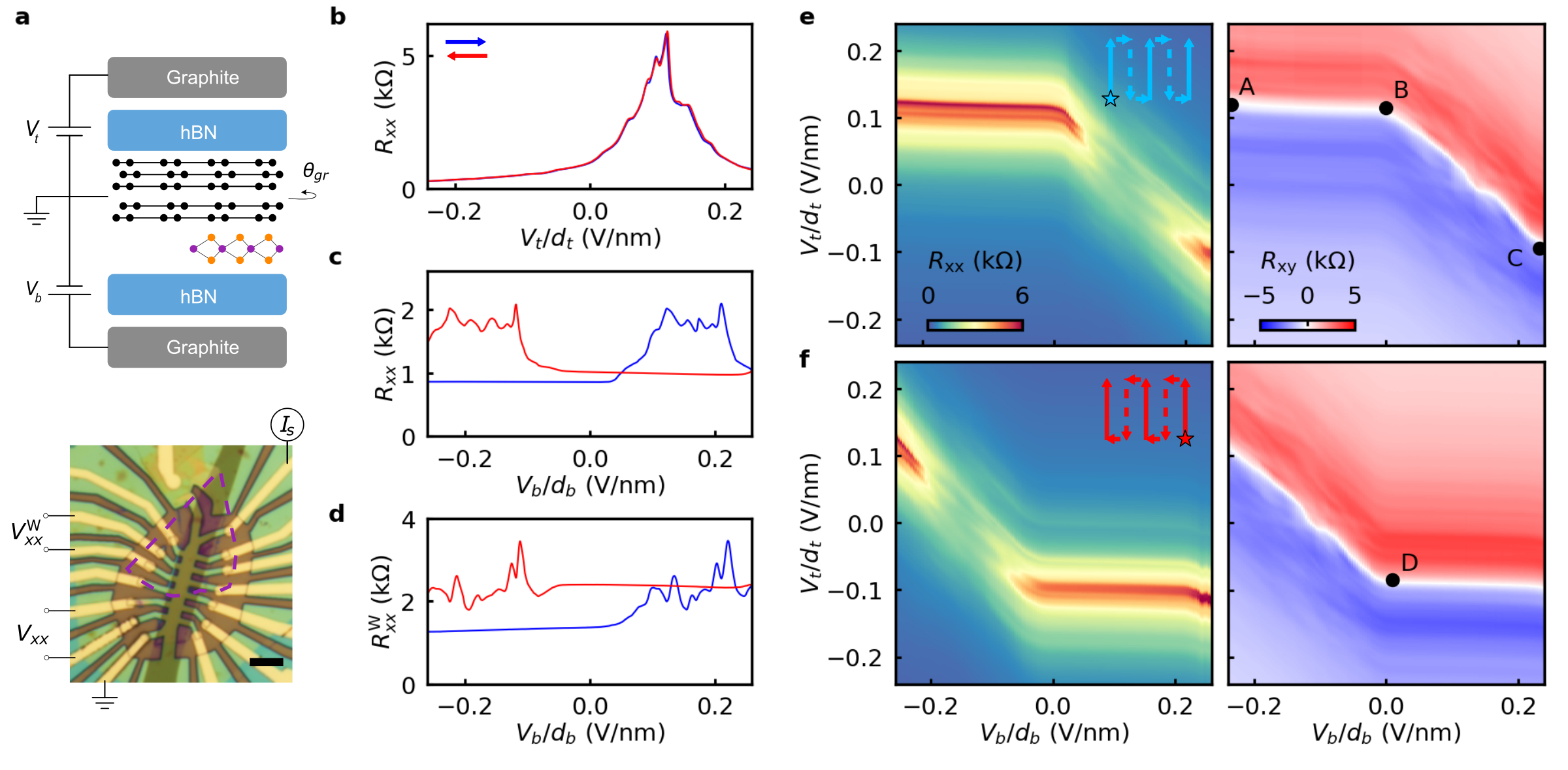} 
\caption{\textbf{Hysteresis in a 5-layer device with a partial WSe$_2$ monolayer spacer.}
\textbf{a}, (top) Cartoon schematic of a 5-layer device, comprising twisted Bernal bilayer and trilayer graphene with a twist angle of $\theta_{\rm gr}=0.65^{\circ}$. A portion of the channel has a monolayer WSe$_2$ spacer between the bilayer graphene and bottom hBN. 
(bottom) Optical micrograph of the device. The purple dashed curve outlines the monolayer WSe$_2$ flake. Measurement setup annotations are similar to Fig. 1a. The resistance across the region without WSe$_2$ is $R_{xx}=V_{xx}/I_s$. The resistance across the region with WSe$_2$ is $R_{xx}^{\mathrm{W}}=V_{xx}^{\mathrm{W}}/I_s$. Analogous voltage probes across the device for the $R_{xy} (R_{xy}^{\rm W})$ measurements are omitted for clarity. Scale bar is $5~\rm \mu m$.
\textbf{b-c}, $R_{xx}$ as a function of top (\textbf{b}) and bottom (\textbf{c}) gate bias in the region of the device without WSe$_2$. In both measurements, the other gate is held at ground.
\textbf{d}, $R^{\rm w}_{\rm xx}$ versus $V_b$ with $V_t=0$ in the region of the device with WSe$_2$.
\textbf{e-f}, Maps of $R_{xx}$ at $B=0$ and $R_{xy}$ at $B=8$~T as a function of both $V_t$ and $V_b$, where $V_b$ is swept from negative to positive (\textbf{e}) or from positive to negative (\textbf{f}). Measurements are shown from the region of the device without WSe$_2$. The maps are taken by sweeping $V_t$ as the fast axis (negative to positive) and $V_b$ as the slow axis. The arrows show the relevant gate sweeping directions, where data is collected only for the sweeps corresponding to solid arrows, and the star represents the starting point of the measurement.
}
\label{fig:2}
\end{figure*}

Prior sightings of the GDW/ratchet effect have mostly been in various bilayer graphene device geometries either presumed or confirmed to have rotational alignment between the graphene and at least one of the hBN flakes~\cite{Zheng2020,Niu2022,zheng2023electronicratcheteffectmoire,Yan2023}. However,the importance of such alignment, and the dependence on the corresponding moir\'e periodicity, remains to be conclusively established. A difficulty in working with bilayer graphene channel devices is the impossibility of treating the two sides of the channel separately, since the bilayer effectively contains a single sheet of carriers whose resistance depends on the total charge density and the penetrating displacement field. This is not the case with our graphite device due to its complete screening of displacement field, and as a result we can unambiguously associate the hysteresis with only the bottom surface of the graphite flake. In addition, to the best of our knowledge this interface is not aligned with hBN (see Methods and Supplementary Information Fig. \ref{ED_fig:BN_alignment}). Combined, these observations make it very unlikely that moir\'e effects at this interface are a significant factor.

\medskip\noindent\textbf{Hysteresis with and without a WSe$_2$ spacer}

We next examine a device with a 5-layer thick channel consisting of Bernal trilayer graphene atop Bernal bilayer graphene with a $0.65^{\circ}$ misalignment between them. Additionally, this device has monolayer WSe$_2$ inserted between the graphene and the bottom hBN over roughly half of the channel (Fig. 2a). Just as in the 24-layer device, the resistance $R_{xx}$ is not hysteretic in top gate voltage $V_t$ (Fig. 2b) but exhibits large hysteresis on sweeping bottom gate voltage $V_b$ (Figs. 2c-d). The traces are again roughly symmetric about $V_b=0$, and over certain intervals in each sweep direction $R_{xx}$ seems almost independent of $V_b$ (GDW effect). Remarkably, almost exactly the same behavior is seen in both parts of the device with and without the monolayer WSe$_2$ insertion (Figs. 2c-d).

In contrast with the 24-layer graphite device considered earlier, the 5-layer graphene channel is too thin to completely screen electric fields. Its resistivity thus depends on linear combinations of the two gate voltages, resulting in a diagonal trajectory of the charge neutrality point in a dual-gate map. The multiple sub-peaks in $R_{xx}$ are caused by the Fermi energy moving through minibands created by the long wavelength moir\'e pattern formed at the bilayer/trilayer graphene interface ~\cite{Waters2024} (see Supplementary Information Fig. \ref{ED_fig:twist_angle_tMM}). In a normal non-hysteretic device, these resistance features would follow parallel diagonal lines in a dual-gate map corresponding to contours of fixed doping, with slopes equal to the capacitance ratio of the two gates. Figures 2e-f show dual-gate maps of $R_{xx}$ (at $B=0$~T) and $R_{xy}$ (at $B=8$~T) for the 5-layer device, measured by sweeping $V_t$ while incrementally changing $V_b$ either forward (Fig. 2e) or backward (Fig. 2f). We identify the charge neutrality point (CNP) as being where $R_{xy}$ changes sign. What is seen here closely resembles the hysteresis behavior reported in bilayer graphene devices. When $V_b$ is increased starting from a negative value (position A in Fig. 2e), the CNP and the $R_{xx}$ features initially seem to be ``stuck'' at nearly constant values of $V_t$ (GDW effect). At point B, the behavior changes and the CNP moves along a more familiar-looking diagonal line up to the highest gate voltage (point C). On subsequently decreasing $V_b$ (Fig. 2f) the CNP appears to be stuck at a different value of $V_t$ (electron ratchet effect). At point D the CNP starts to move diagonally again. Overall, the CNP traces out a parallelogram-like loop. Careful analysis of the 24-layer device reveals essentially the same underlying behavior (see Methods and Supplementary Information Figs. \ref{ED_fig:Drude_model}-\ref{ED_fig:tMM_B_0p5T_maps}).

\begin{figure*}[t]
\includegraphics[width=\textwidth]{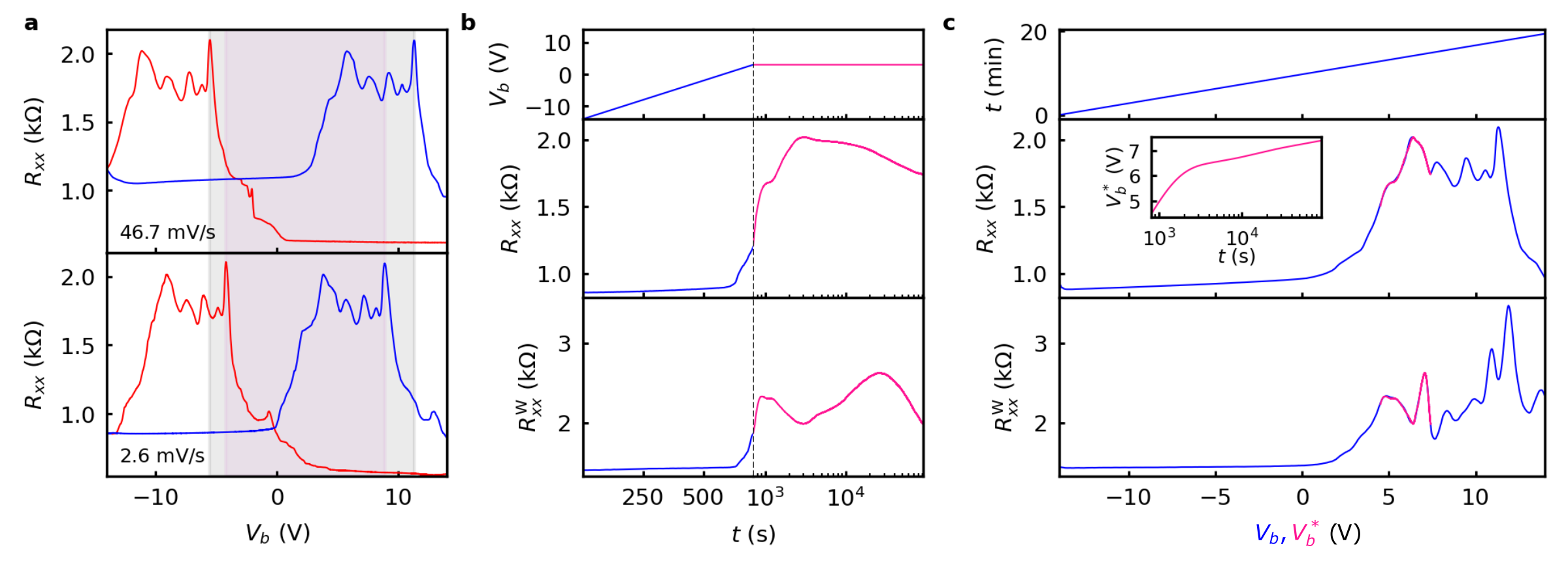} 
\caption{\textbf{Analysis of charge relaxation timescales in the 5-layer device.}
\textbf{a}, (top) Measurement of $R_{xx}$ in the region of the 5-layer device without the WSe$_2$ spacer as $V_b$ is swept from negative to positive (blue) and positive to negative (red) at a constant rate, $dV_b/dt = 46.7~\rm mV/s$, with each sweep taking a total of approximately 10 minutes. (bottom) Same measurement, but at a slower sweep rate, $dV_b/dt = 2.6~\rm mV/s$, with each sweep taking a total of approximately 3 hours. The gray and pink shadings show the offset of a common peak in each measurement.
\textbf{b}, Measurements of $R_{xx}$ ($R_{xx}^{\rm W}$) in the region of the device without (middle) and with (bottom) WSe$_2$ as a function of time. The top panel shows $V_b$ as a function of time, with fixed $V_t=0$. The vertical black dashed line indicates the time at which both gates are held at fixed values. 
\textbf{c}, Blue curves show measurements of $R_{xx}$ ($R_{xx}^{\rm W}$) in the region of the device without (middle) and with (bottom) WSe$_2$ as $V_b$ is swept from negative to positive. The top panel shows the evolution of $V_b$ with time. The pink curves are the same data from the corresponding panels in \textbf{b} converted from time to $V_b^*$ using the non-linear conversion shown in the inset.
}
\label{fig:3}
\end{figure*}

\medskip\noindent\textbf{Charge relaxation timescale}

Although the GDW/ratchet effect appears to be quite reproducible and insensitive to gate sweep speed on typical timescales of minutes-long sweeps, differences can be seen over longer time scales. Figure 3a compares measurements of $R_{xx}$ in the 5-layer device for a cycle of $V_b$ performed over 10 minutes ($dV_b/dt = 46.7~\rm mV/s$, top) and 3 hours ($dV_b/dt = 2.6~\rm mV/s$, bottom). The hysteresis is slightly diminished in the slower measurement. To get quantitative information about relaxation timescales, we quickly sweep $V_b$ at $V_t=0$, from its most negative value to a small positive value close to where the CNP appears to start moving (point B in Fig. 2f), then measure $R_{xx}$ at fixed voltage as a function of time, $t$, for 48 hours (Fig. 3b). The resulting trace is nonmonotonic and reminiscent of a stretched gate sweep. 

Moreover, the detailed features in $R_{xx}(t)$ can be made to coincide perfectly with those in a fast measurement of $R_{xx}(V_b)$ by suitable choice of a map from $t$ to $V_b$, as shown in Fig. 3c. We used a map of the form $V^*_b=V_0 + \sum_i V_i e^{(-t/\tau_i)}$, where $\tau_i$ characterize the timescales of charge relaxation and $i=1,2,3...$. Three timescales are needed for the data shown in Fig. 3c: ($6.60\times 10^2, 1.29\times 10^4, 3.30\times 10^5$) s. The procedure is discussed further in the Methods, but we emphasize that although a non-linear mapping with many parameters is required, the fact that there exists a conversion from $R_{xx}(V_b)$ to $R_{xx}(V_b^*(t))$ is highly suggestive that long-time scale charge dynamics are at play in the system. We further find that such a mapping could be performed for all values of $V_b$ we have held at for long periods of time (as well as other parameters such as $V_t$, see Supplementary Fig. \ref{ED_fig:timescales}). We conclude that the charge configuration of the devices lies in a continuum of history-dependent states with very long relaxation time scales. Furthermore, we find the fitting parameters to be the same for the measurements on the device regions with and without the monolayer WSe$_2$, which we discuss further in the Methods.

\medskip\noindent\textbf{Consideration of the underlying mechanism}

Explanations for the GDW/ratchet phenomenon can be divided into two types: either the charge that accumulates due to the changing gate voltage is localized in the dielectric, or the charge accumulates within the graphene itself but for some reason is not mobile. The latter requires a novel strongly correlated electronic mechanism that allows a fraction of the electrons in the channel to be localized while others remain itinerant. For the case of bilayer graphene aligned with hBN, it has been proposed that the long-wavelength moir\'e pattern localizes carriers in one graphene layer owing to formation of a dipolar exciton condensate, with carriers in the remote layer remaining itinerant~\cite{zheng2023electronicratcheteffectmoire}. We can exclude this as a possibility in our devices, since the hysteresis is associated with an hBN flake that is either far from alignment with the graphene or separated from it by monolayer WSe$_2$, precluding the formation of the required moir\'e pattern. Additionally, the stronger screening in thicker graphite flakes should suppress correlation effects relative to bilayers, and the persistence of the phenomenon to room temperature requires an implausibly large correlation energy scale ($>>25$~meV).

We now consider the former, more conventional, explanation involving charge redistribution in the dielectric. Since in the 24-layer device the hysteresis is purely associated with the lower graphite surface and bottom gate, we will only consider the electric fields and charges on one side (the bottom) of the channel. We assume a parallel-plate capacitor model with a distance $d$ between the gate and channel, whose bottom surface is at position $x=0$, with a distributed volume free charge density $\rho(x)$ present in the dielectric. For now, we do not discuss the nature of $\rho(x)$. Figure 4a shows an electron energy diagram of this scenario, with $\rho(x)$ in blue and the electrostatic energy, $-e \varphi(x)$, where $e$ is the charge of the electron, shown in red (see Methods for a detailed analysis).

\begin{figure*}[t]
\includegraphics[width=\textwidth]{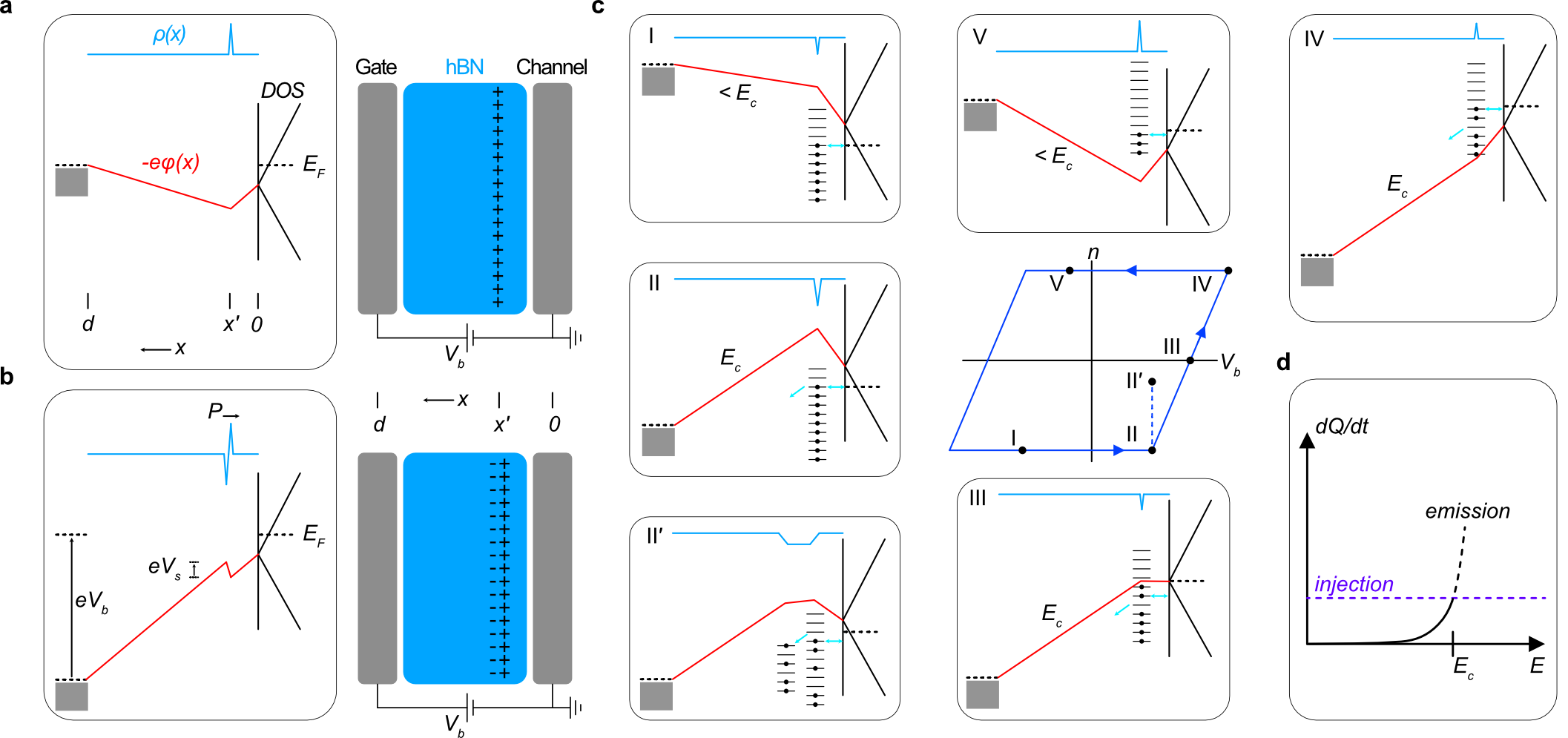} 
\caption{\textbf{Phenomenological scenarios for modeling the GDW/ratchet effect.} 
\textbf{a}, (left) Electron energy diagram showing the effect of a large free charge concentration ($\rho(x)$, blue) confined to a single internal sheet of hBN. The red line shows the corresponding electrostatic energy profile ($-e\varphi(x)$) across the dielectric. The right-hand side of the diagram shows the density of states (DOS) of the channel (assumed here to be monolayer graphene for simplicity) and the Fermi energy, $E_F$. The channel and gate are both held at ground. (Right) Device schematic illustrating the sheet of free charge in the hBN. 
\textbf{b}, Same as (\textbf{a}), but for a sheet of dipoles formed by adjacent hBN layers with AA-type ferroelectric stacking. The channel is grounded and the gate is held at $V_b$. $P$ is the polarization density and $V_s$ is the potential across the AA-stacked interface (see Methods).
\textbf{c}, (central panel) Schematic of the parallelogram in ($V_b, n$) space that is traced out when sweeping the hysteretic gate. (outer panels) Electron energy diagrams, labeled I-V and II$'$ corresponding to the respective points on the parallelogram, following a counter-clockwise evolution around the loop from I to V. II$'$ is accessed by stopping at II and waiting without changing $V_b$. In the electron energy diagrams, there is assumed to be a band of defect states near the hBN surface adjacent to the channel. Horizontal blue arrows depict electrons tunneling between defect states and the channel. In II-III-IV, diagonal blue arrows denote charges that are field-emitted into the hBN conduction band and travel to the gate. In II$'$ diagonal blue arrows denote charge equilibration between the near-surface states in the hBN and defect states deeper within the dielectric.
\textbf{d}, Schematic of the rate of charge flow, $dQ/dt$, as a function of electric field, $E$, in and out of the near-surface states. The critical field, $E_c$, is reached when the rate of charge injection from the channel equals the rate of field emission from those states towards the gate.
}
\label{fig:4}
\end{figure*}

The existence of hysteresis implies that $\rho(x)$ has multiple configurations at a given $V_b$. The short timescales and high reproducibility of the GDW effect rule out ionic drift and structural changes, with one possible exception: it has recently been shown that adjacent layers of hBN with AA-type stacking can produce ferroelectricity~\cite{Yasuda2021}. An interface with such stacking within the hBN could thus carry an electric polarization that reverses polarity in response to a changing gate electric field, caused by the hBN layers sliding with respect to each other by a fraction of a lattice constant ~\cite{Yasuda2021, Vizner2021-vdWsliding, Wang2022TMDs, Weston2022_marginallytwistedMoS2, ding2024}. For a thin sheet of areal polarization density $P$ located within the hBN (Fig.~\ref{fig:4}b) we estimate a maximum associated hysteresis in $V_b$ of $\approx100$~mV for full polarization reversal of the interface. Domain effects will only reduce the net ferroelectric polarization. This is at least an order of magnitude smaller than the hysteresis observed in our devices, and also corresponds to the wrong sign of the gate screening. Sliding ferroelectricity at a single hBN interface can thus be ruled out as the root cause of the phenomenon (see Methods for additional analysis). 

The remaining possibility is that $\rho(x)$ changes because electrons move between locations in the hBN. To analyze this scenario, we first summarize the observed phenomenology by way of the schematic in Fig. 4c. The central blue lines show the trajectory of the graphene doping as $V_b$ is swept with fixed $V_t=0$. The doping traces out a characteristic parallelogram, seen regularly in bilayer graphene devices~\cite{Zheng2020,Niu2022} or the 5-layer device in this work (Fig. \ref{fig:2} and Supplementary Information Fig. \ref{ED_fig:Hall_density}). We next draw inferred electron energy diagram cartoons for key points on the parallelogram (Fig. 4c, I-V). For simplicity we only consider $\rho(x)$ to be non-zero near the channel, since charge in the hBN near the gate does not have a significant influence on the doping in the channel. Within this framework, the GDW trajectory I-II can only be explained if there is a peak in $\rho(x)$ near the hBN surface, where negative screening charge accumulates as $V_b$ increases. At point II this charge accumulation stops, and thereafter $n$ changes rapidly with $V_b$ (along II-III-IV). When the sweep direction is reversed at IV, the GDW trajectory IV-V can only be explained by positive charge accumulation near the surface of the channel. Analogous reasoning can be followed around the negative-$V_b$ side of the hysteresis loop.

The cartoons in Fig. 4c also include the ingredients of a model that can qualitatively explain this behavior. Along I-II, the evolution of $\rho(x)$ implies the presence of electron states near the surface of the hBN next to the channel. Such states must be close to the electrochemical potential, $\mu$, in the channel. Their timescale for exchanging electrons with the channel must be much shorter than the typical gate sweeping time, and their density of states must be much larger than that of the channel for them to effectively screen the back gate (the GDW effect). The analogous behavior along IV-V (and, correspondingly, at negative $V_b$) implies that the near-surface states can have net positive or negative charge, i.e, that both donors and acceptors (or, possibly, negative-U centers) are present. 

The halting of the screening (i.e., GDW) behavior at II can be explained as follows: at this point the electric field (the slope of the red line in the cartoons) in the middle of the hBN reaches a critical value, $E_c$, such that electrons in the near-surface states can be field-emitted into the hBN conduction band and drift to the gate. Since the field-emission rate typically grows exponentially with electric field, e.g. in the Poole-Frenkel effect ~\cite{Frenkel1938}, whereas we assume a fixed rate of charge injection from the channel to the near-surface states, the internal electric field in the hBN cannot exceed $E_c$ even upon further increasing $V_b$ (Fig.~\ref{fig:4}d). Instead, as $V_b$ increases (II-III-IV), the near-surface--state charge decreases self-consistently in such a way as to maintain the bulk electric field close to $E_c$. The near-surface--state population is in a dynamical state in this regime, with a flow of electrons through the hBN, that results in effective anti-screening of the channel (see Methods). However, as soon as the sweep direction is reversed (at IV), the bulk electric field falls below $E_c$ and the surface states once again come into equilibrium with the channel, screening it from further changes in $V_b$ (the ratchet effect).

Slow relaxation of the neutrality point, such as along the trajectory I-II’, implies that when the field in the bulk is close to $E_c$ it is possible for the charge in the near-surface states to change with time even while they remain nearly in equilibrium with the channel. This can be explained by the presence of defect states deeper in the hBN bulk, which capture or release electrons on longer timescales. As a result, the peak in $\rho(x)$ will tend to spread into the hBN over time, through a continuum of configurations, and the doping in the channel will consequently decrease with time at fixed gate voltage, consistent with our observations in the 5-layer device (Fig. \ref{fig:3} and Supplementary Information Fig. \ref{ED_fig:timescales}).

\medskip\noindent\textbf{Discussion}

Despite the new constraints our findings place on the origin of the GDW/ratchet behavior, crucial open questions remain. Most significantly, we have not been able to identify the putative near-surface states in the hBN proposed within our mobile-charge model. We have studied devices with one hBN dielectric having a large density of carbon substitutional defects~\cite{CarbonRichBN}, as well as devices with one hBN that was neutron irradiated (thought to create a large density of boron vacancies and possible lithium substitutions~\cite{Li2021_neutronirradiatedBN}). The other hBN in these devices was a typical, nominally pristine flake. The neutron-irradiated hBN devices exhibit weak hysteresis upon sweeping the gate with the defect-rich dielectric, but do not show GDW/ratchet behavior (see Supplementary Information Figs. \ref{ED_fig:tMM_carbon-defect-devices}-\ref{ED_fig:tMM_IrBN-devices}). Thus, if defects are the origin of the latter effect, they must be some other type than those we have studied so far. Another possibility is that a layer of mobile charge is somehow created at the hBN surface due to some form of electrical or mechanical stress created upon stacking the vdW heterostructure and first operating the gate. New characterization techniques would be helpful for probing such effects, especially probes that can provide detailed information about the microscopics of the graphene/hBN interface.

Another key open question is whether a moir\'e pattern is strictly necessary for generating the GDW/ratchet effect. There is a considerable body of evidence that it may indeed be needed in bilayer graphene devices, although the situation is not completely clear as it is not always possible to identify the interfacial twist angle unambiguously. In contrast, our thicker devices do not appear to have alignment between the hysteretic graphene/hBN interface, although we do not have a precise measure of the twist angles of these interfaces, and there are other known interfaces within our devices that do have a long-wavelength moir\'e. We believe these do not play a role owing to various screening considerations, but cannot conclusively rule out their influence being important. Additionally, the persistence of the effect with a WSe$_2$ monolayer insertion in our 5-layer device, along with reports of similar effects in a device with an artificially stacked MoS$_2$~\cite{niu2024tMoS2} substrate and various forms of hysteresis seen in graphene devices on CrI$_3$ and RuCl$_3$,~\cite{Rossi2023, Zhou2019_GrRuCl3, Tseng2022} raises questions about whether other vdW dielectrics can support similar GDW/ratchet behavior. Our observations suggest that it could be easier to realize than previously thought, and point to new directions for exploring and eventually harnessing this intriguing behavior.

\section*{Methods}

\textbf{Device fabrication.} Flakes of graphite and hBN were exfoliated onto silicon/silicon dioxide wafers and identified using optical microscopy. The exofoliated graphene multilayers were separated into two distinct regions using polymer-free anodic oxidation nanolithography~\cite{Li2018}. For the 5-layer device, this involved identifying a flake with a connected bilayer and trilayer region. For the 24-layer device, this involved cutting a 12-layer flake into two separate pieces. The vdW heterostructures were sequentially assembled using a polycarbonate (PC)/polydimethylsiloxane (PDMS) stamp in the following order: graphite, hBN, twisted graphene, hBN, graphite. The graphene flakes were rotationally misaligned by rotating the stage after picking up the first graphene sheet. For the 5-layer device, a monolayer WSe$_2$ was additionally picked up after the twisted graphene, and covered only a portion of the channel. After assembly, the completed vdW stack is dropped onto a clean Si/SiO$_2$ wafer. We used standard electron beam lithography and CHF$_3$/O$_2$ plasma etching to define vdW stacks into a Hall bar geometry and standard metal deposition techniques (Cr/Au) to make electrical contact ~\cite{Wang2013}.

In the 24-layer device, the top and bottom hBN thicknesses are $d_t=51\pm4$~nm and $d_b=57\pm3$~nm. For the 5-layer device, they are $d_t=26\pm3$~nm and $d_b=58\pm2$~nm. These values are determined by tapping mode atomic-force microscopy (AFM). Reported values are averages calculated by measuring at different spots within the device, rounded to the nearest integer. Uncertainties are the calculated standard deviations from these measurements, rounded up to the nearest integer.

\textbf{Transport measurements.} Transport measurements on the 5- and 24-layer graphene devices were carried out in a Cryomagnetics variable temperature insert (base temperature $1.7 ~\rm K$; $350 ~\rm mK$ with a $^3$He insert) using a lock-in amplifier with frequencies between 13.33 and 17.77 Hz and a.c. bias between $10$ and $200$~nA. All measurements were performed at $1.7 ~\rm K$ unless noted otherwise. Additional transport measurements on these devices, and devices built with carbon-rich hBN and neutron irradiated h$^{10}$B$^{15}$N dielectrics were carried out in a $^4$He cryostat (base tempearture $4$~K) using a lock-in amplifier with frequencies between 13.33 and 17.77 Hz and an a.c. bias between 10 and 500 nA.

We denote all of the longitudinal resistance measurements by $R_{xx}$ throughout the manuscript. Although the 2D sheet resistivity, $\rho_{xx}$, is not well-defined for the 24-layer device owing to its bulk nature, both the 24 layer and 5 layer devices have geometries such that the channel aspect ratio $W/L\approx 1$, where $W$ ($L$) is the channel width (length).

\textbf{Extraction of twist angles.} We estimate the twist angle, $\theta_{gr}$, between graphene flakes by fitting the sequence of Brown-Zak (BZ) oscillations that emerge upon application of a magnetic field. BZ oscillations occur when a rational fraction of magnetic flux quanta penetrate the \moire unit cell, $\phi/\phi_0=(4B/n_s)/\phi_0=p/q$, where $\phi_0=h/e$, $h$ is Planck's constant,  and $p$ and $q$ are integers. The superlattice density, $n_s$, can be extracted by fitting the sequence of conductance peaks as $B$ is increased (Supplementary Information Fig. \ref{ED_fig:twist_angle_tMM}). The superlattice transport features in both devices arise due to the twisted graphene interface, rather than a \moire pattern formed at the graphene/hBN interface (as discussed below). As such, we use $n_s=8\theta_{gr}^2/\sqrt{3}a^2$ to determine the twist angle, where $a=0.246~\rm nm$ is the graphene lattice constant, and find that $\theta_{gr}=1.21^{\circ}$ in the 24-layer device and $\theta_{gr}=0.65^{\circ}$ in the 5-layer device. Owing to the small twist angle in the 5-layer device, several \moire superlattice features are observed when tuning the density at low magnetic field. We can measure the spacing of these features when sweeping the non-hysteretic gate voltage, $\Delta V_t$, and see that it is consistent with the twist angle extracted from the BZ oscillations, i.e. that $C_t\Delta V_t/e=n_s$.

There is no signature of Hofstadter butterfly features indicative of a graphene/hBN moir\'e potential in either device. Instead, we estimate the twist angles by inspecting straight crystalline edges of the graphene and hBN flakes (optical micrographs in Supplementary Information Fig.~1). Since we did not perform Raman spectroscopy or second harmonic generation measurements before device fabrication, we cannot distinguish zig-zag and armchair crystal edges and therefore can can only obtain the graphene/hBN alignment modulo $30^{\circ}$. In the 24-layer device, we find that the twist angle between the top graphite flake and the top hBN is $\theta_t = 2^{\circ} (\mathrm{mod}\ 30^{\circ})$. The twist angle between the bottom graphite flake and the bottom hBN is $\theta_b = 5^{\circ} (\mathrm{mod}\ 30^{\circ})$. For the 5-layer device, we find $\theta_t = 0.5^{\circ} (\mathrm{mod}\ 30^{\circ})$ and  $\theta_b = 6.5^{\circ} (\mathrm{mod}\ 30^{\circ})$. We estimate an uncertainty in the optically determined twist angle to be $\approx1^\circ$, however, this does not take into account the uncertainty in properly identifying perfectly straight edges with pristine armchair or zigzag termination purely from optical micrographs. 

\textbf{Calculation of the doping offset.} The offset in doping is calculated using $n_{\rm hyst}= 2C_b \Delta V_b/e$ where $C_b=\varepsilon_r \varepsilon_0/d_b$ is the areal capacitance, and $\varepsilon_r=3$ is the dielectric constant of thick hBN.

\textbf{Analysis of the device electrostatics.} Here, we derive the basic electrostatics underlying the band diagrams shown in Fig. 4. For simplicity we will assume uniformity in the plane of the device, and thus a one-dimensional problem: this is justified by the sharpness of the moiré-miniband resistance peaks in the 5-layer device. Let the surface of the graphite channel be $x=0$ and the surface of the gate electrode surface be $x=d$, forming a parallel-plate capacitor with plate separation $d$. Keeping in mind the 24-layer device, we assume that negligible electric field penetrates through either the gate electrode or the thin graphite channel. Applying Gauss’s law then shows that a sheet of (positive) free charge of area density $\sigma_0$ at position $x=x'$ in the hBN induces screening area charge densities of $-(1-x'/d) \sigma_0$ in the channel and $(-x'/d) \sigma_0$ in the gate. Note that this is independent of the dielectric constant, $\varepsilon_r$, and that the values of $d$ and $x'$ can include effects of finite screening lengths. Naturally, the closer the free charge is to the channel (i.e., the smaller is $x'$), the larger the fraction of the screening charge residing in the channel. The cartoon in Fig. 4a indicates the volume free charge density $\rho(x)=\sigma_0 \delta(x-x')$ (blue) and the electrostatic electron energy $-e\phi(x)$ (red) at $V_b=0$. The latter obeys $d^2\phi/dx^2 = -\rho/(\varepsilon_r \varepsilon_0)$, and is drawn relative to the graphene Dirac point, which is below the electrochemical potential $\mu$ due to the accumulation of electrons in the channel.

By linearity, a general distributed volume free charge density $\rho(x)$ in the hBN induces an areal charge density $-\int_0^d (1-x/d)\rho(x) dx$ in the channel surface. If the channel is grounded and there is a voltage $V_b$ on the gate, then by superposition the net electron doping of the channel is 
\begin{equation}
n=\frac{1}{ed} \left(\varepsilon_r \varepsilon_0 V_b+\int_0^d (x-d)\rho(x) dx\right).
\end{equation}
Here, the first term is the usual capacitive contribution, $n=C_b V_b/e$. Note that if the work function difference between the channel and gate is not negligible it simply provides an offset on $V_b$. 

\textbf{Estimate of sliding ferroelectricity hysteresis.} For a thin sheet of areal polarization density $P$ located within the hBN, such as might be associated with an AA' stacking fault, the integral in Eq.~1 is just equal to $P$. Since $P\approx \varepsilon_r \varepsilon_0 V_s$, where $|V_s|\approx50$~mV is the potential drop across the sliding interface~\cite{Wang2022TMDs}, Eq. (1) gives $n\approx \frac{\varepsilon_r \varepsilon_0}{ed}(V_b+V_s)$. Thus, $V_s$ effectively adds to the gate voltage, as illustrated in Fig. 4b. Hence, the maximum possible hysteresis in $V_b$, corresponding to full polarization reversal of the interface, is $2V_s \approx 100$~mV, much smaller than what is observed. In principle, the magnitude of the ferroelectric effect could be increased by having many AA' interfaces within the hBN. However, not only is this a very unlikely scenario given the correlation of the behavior with other factors in our experiments, but also the GDW effect is exactly the opposite of what one would expect to result from such ferroelectricity. An applied electric field will tend to cause ferroelectric polarization within the hBN to flip in a sense that it increases the electric field at the graphene, thereby enhancing the resulting channel doping.  

\textbf{Anti-screening behavior.} Consider a change in $V_b$ of $\delta V_b$, causing a change in surface-layer charge $\delta n_s$, and change in doping $\delta n$. To maintain the electric field at $E_c$ in the middle of the hBN, the potential change $\delta V_b$ must be dropped between the surface layer(s) and the channel or gate (on either side of the hBN). This changes the electric field at the graphene by $\approx \delta V_b/d_0$ where $d_0$ is an effective distance of the order of the thickness of hBN where the electric field is smaller than $E_c$, leading to $\delta n \approx \varepsilon_r \varepsilon_0 \delta V_b/d_0$. 

This analysis predicts that the slope of the CNP in II-III-IV region of Fig.~\ref{fig:4}c should be larger than predicted from the ratio of the gate capacitances by a factor $d/d_0$. However, we do not see this behavior experimentally, within our ability to resolve such an effect. A detailed measure of this deviation would require accurate knowledge of the capacitance of each dielectric. The value of the capacitance is usually determined by fitting the slopes of quantum oscillation features or the Hall resistance upon applying a magnetic field. However, an accurate measure of the hysteretic gate capacitance through these methods is precluded by the GDW/electron-ratchet effects and the long charging timescales. Therefore, we must rely on estimating the capacitance by measuring the thickness of the hBN flakes in AFM and assuming the dielectric constant of hBN. The slope of the CNP in II-IV of Fig.~\ref{fig:4}c is equivalent to the capacitance ratio to within our experimental uncertainty. 

\textbf{Mapping from $t$ to $V_b$ in long-hold measurements.} 
To characterize the long-timescale charging dynamics in these systems, we first take a standard measurement where we ramp $V_b$ in one direction over the entire accessible gate range of the device while holding $V_t$ fixed. Next, we repeat the measurement, sweeping $V_b$ in the same direction, but stopping at a value $V_b^{h}$, and holding the gate voltages fixed over the course of many hours or days while recording the resistance continuously. The two measurements yield resistance as a function of voltage, $R_{xx}(V_b)$, and resistance as a function of time, $R_{xx}(t)$, respectively. As noted in the main text, we can then map features of $R_{xx}(t)$ to $R_{xx}(V_b)$ with a non-linear scaling of $t$ according to the equation
\begin{equation}
V^*_b=V_0 + \sum_i V_i e^{(-t/\tau_i)},
\end{equation}
such that we can perform a least-squares fit between $R_{xx}(V_b^*(t))$ and  $R_{xx}(V_b)$ to find the parameters $V_i$ and $\tau_i$. We can perform the same fit to acquire the parameters for the region of the device with the WSe$_2$ spacer as well,  $V_i^{\rm W}$ and $\tau_i^{\rm W}$.

We choose an exponential mapping because a simple linear mapping from $t$ to $V_b^*$ does not fit the data. A single exponential time scale is also insufficient, so we include multiple exponential terms in the fit procedure. The number of terms we choose depends on how long we performed the hold measurement. The resistance continuously evolves no matter how long we hold the gate voltages fixed, but the changes in resistance occur on progressively longer timescales as we wait. We find that by performing hold measurements up to 48 hours ($\approx 2\times 10^5 ~\rm s$), we can probe characteristic relaxation timescales up to $\tau \approx 10^5~\rm s$. We exclude the hold data for $t<10~\rm s$ from our fits, such that we are not considering ``conventional'' measurement hysteresis effects arising on short timescales (e.g., owing to the integration time of the lockin amplifier). For measurements in which we start the hold in the GDW regime, or at the onset of normal field effect behavior, there is a significant initial period of time in which the resistance essentially does not change, before subsequently evolving noticeably with longer wait times. In such cases, we also exclude the first $t\approx10 - 100$~s of the measurement from our fitting, as these cannot be matched to features in the corresponding gate sweep.

We perform fits for several different measurement configurations. Fig.~\ref{fig:3}b-c show the case of $V_b^h=+3~\rm V$ and $V_t=0~\rm V$. Supplementary Information Fig.~\ref{ED_fig:timescales} shows similar measurements for several other starting conditions, including $V_b^h=0~\rm V$ and $V_t=0~\rm V$ but initially sweeping in the opposite direction prior to the hold, and hold measurements with $V_t\neq 0$. We report the resulting fit parameters in Supplementary Information Table~S1. Given the large number of fitting parameters, we do not try to infer anything quantitative about the dependence of the relaxation dynamics on the gate voltages. Instead, we highlight that we are able to successfully perform such a fitting procedure for all the measurement configurations we attempted, indicating that long-timescale charge relaxations are ubiquitous features of the device behavior. All the data in Table~S1 are for the 5-layer device. We also performed a hold measurement in the 24-layer device  and see similar relaxation dynamics (Supplementary Information Fig.~\ref{ED_fig:tMM_hold_measurements}), but we are unable to extract quantitative relaxation timescales for that device since the measurement is performed over a comparatively shorter waiting time. Finally, we note that we do not report the value of $V_0$ for each fit. This value should be close to the hold voltage, i.e. $V_0 \approx V_b^h$, however, the onset of ``normal'' field effect (or the extent of the GDW effect) has some variation and dependence on measurement conditions (e.g., Fig.~\ref{fig:3}a and Supplementary Information Fig.~\ref{ED_fig:tMM_sweep_ranges}-\ref{ED_fig:reproducibility}). Thus, in practice, $V_0$ can be significantly different than $V_b^h$ and does not hold physical significance. 

\textbf{Tunneling rate with monolayer WSe$_2$.} In Fig.~\ref{fig:3} and Supplementary Information Fig.~\ref{ED_fig:timescales}, we show that the exact same mapping between $t$ and $V_b$ can be used to fit the data in the device regions with and without the WSe$_2$ spacer. This is established by comparing the charge relaxation timescales and finding them to be the same for the two regions, $\tau_i=\tau_i^{\rm W}$. Further, we argue that the mechanism of the charge relaxation/GDW effect requires tunneling between the channel and defect states in the hBN (Fig. \ref{fig:4}c). Naively, this should imply different charge relaxation timescales between the regions of the device with and without the WSe$_2$ spacer, i.e. $\tau_i<\tau_i^{\rm W}$, due to the additional tunnel barrier in the latter. However, since there is a lateral junction between the two regions of the device (Fig. \ref{fig:2}a), it is possible for charges to drift within the channel to the region of the device without the WSe$_2$ spacer. From there, these charges can tunnel into the hBN, thus effectively bypassing the additional WSe$_2$ barrier. The timescale for charge transport across the lateral junction is likely much shorter than the long-timescale charging associated with the hBN near-surface states. Such a scenario is consistent with our observation that the extracted charge relaxation timescales are the same for the region of the device with and without WSe$_2$, $\tau_i=\tau_i^{\rm W}$. Nevertheless, we stress that the observation of the GDW/ratchet effects in the WSe$_2$ spacer region, which are seen even when the source and drain electrodes are entirely within this region (Supplementary Information Fig.~\ref{ED_fig:contact_comparison}), indicate that the graphene channel and hBN do not need to be directly adjacent.

\textbf{Equivalence of the GDW effect in the 24-layer and 5-layer devices.}
The dual gate maps in the 5-layer devices show a parallelogram-like pattern (Fig.~\ref{fig:2}e \& f, Supplementary Fig.~\ref{ED_fig:Hall_density}). However, the dual gate maps in Fig.~\ref{fig:1}e show a cross-like pattern for the 24-layer device. Here we seek to show that the two observed patterns have the same origin, where the difference is due to the distinct transport characteristics of two-dimensional graphene and bulk graphite systems. 

In bulk graphite, the top (bottom) gate controls the density of a charge accumulation layer residing on the top (bottom) graphene surface. Deep in the bulk, there are coexisting species of free electrons and holes that are screened from the gates by the surfaces, and thus are not responsive to changing gate voltages. We can utilize a simple multi-component Drude model, adapted from Ref.~\cite{Waters_Nature_2023}, to understand the resulting transport signatures. Two components represent the bulk electron and hole densities, which are equal in magnitude and insensitive to applied gate voltages. Two additional components represent the top and bottom surface charge densities, which are tuned by the gate voltages, $n_t(V_t)$ and $n_b(V_b)$, respectively. The total conductivity is then $\sigma = \sum_i \sigma_i$, where the contribution from each component is given by
$$\sigma_i = \frac{en_i \mu_i}{1+(\mu_i B)^2 }\begin{pmatrix}
1 & \eta_i \mu_i B \\
-\eta_i \mu_i B & 1 
\end{pmatrix},$$
with $\eta_i=\mp 1$ for electrons and holes, respectively, $n_i$ is the density, and $\mu_i$ is the mobility. The resistance tensor is $R=\sigma^{-1}$, with the diagonal (off-diagonal) components representing $R_{xx}$ ($R_{xy}$). 

We calculate the resistance using the same parameters as in Ref.~\cite{Waters_Nature_2023}, but include an additional phenomenological feature to capture the GDW effect. The normal gate, $V_t$, controls $n_t$ according to the usual relation $n_t=C_tV_t/e$. To represent a typical forward sweep of the hysteretic gate, we fix the density on the bottom surface at a value $n_b=n_b^0$ while $V_b$ is less than a critical voltage, $V_{cr}$. When $V_b > V_{cr}$, $n_b$ starts to follow $n_b=n_b^0+C_b(V_b-V_{cr})/e$. The results are shown in Supplementary Fig.~\ref{ED_fig:Drude_model}.

At zero magnetic field, the transport features are primarily cross-like, reminiscent of the data shown in Fig.~\ref{fig:1}e, where the resistance maxima occur when the density on the top and/or bottom surface is near zero. When the magnetic field is turned on, the resistance is maximum when the total charge is zero, which occurs along a diagonal slope in the normal field effect regime, but along a horizontal line in the GDW effect regime, yielding half of the parallelogram-like behavior observed in the two-dimensional devices. Therefore, at finite magnetic field, we expect the bulk graphite device to also exhibit the parallelogram-like behavior when the hysteretic gate is swept forward and backward. We show that this is indeed the case for our 24-layer device at $B=0.5~\rm T$ in Supplementary Fig.~\ref{ED_fig:tMM_B_0p5T_maps}. 

\textbf{Devices with carbon-rich hBN dielectrics.} For preparing devices with carbon-rich hBN dielectrics, doped hBN crystals were first synthesized following the method described in Ref.~\onlinecite{CarbonRichBN}. Secondary ion mass spectroscopy measurements were performed to estimate the concentration of carbon defects in hBN prepared following this technique, yielding values of $\approx10^{18}-10^{19}$~cm$^{-2}$ depending on the specific crystal tested. 

Single-gated devices were constructed using the carbon-doped hBN crystals. The device fabrication procedure was the same as that used for the 5- and 24-layer devices. Devices consist of a graphite channel, and were stacked in the following order: graphite channel, carbon-rich hBN, graphite back gate. The hBN thicknesses in the two devices shown (Supplementary Information Fig.~\ref{ED_fig:tMM_carbon-defect-devices}a-b) are $37$~nm (C-hBN-1) and $20$~nm (C-hBN-2), respectively, as determined by tapping mode atomic-force microscopy. No intentional alignment between the graphite and hBN layers was performed during the stacking process. We do not observe any signatures of hysteresis in these devices (Supplementary Information Fig.~\ref{ED_fig:tMM_carbon-defect-devices}a-b).

\textbf{Devices with neutron-irradiated hBN dielectrics.} For preparing devices with neutron-irradiated hBN dielectrics, high quality h$^{10}$B$^{15}$N crystal flakes were first precipitated from a 48 wt\% nickel, 48\% chromium, and 4 wt\% isotopically enriched boron-10 (99.8\% purity, 98\% isotopic purity flux by the atmospheric pressure, high temperature (ATHP) metal solvent method. The source powders were melted at 1550~$^{\circ}$C and left dwelling for 24 hours before slowly cooling to room temperature. In lieu of natural nitrogen during the crystal growth process, isotopically enriched nitrogen-15 (99.999\% purity, 97\% isotopic purity was used~\cite{Janzen2023}. The resulting crystals were exfoliated mechanically from the metal ingot with thermal release tape. 

The h$^{10}$B$^{15}$N flakes were neutron irradiated at the Ohio State University nuclear reactor for 60~h or 10~h at a power of 300~kW, producing a total fluence of $8.6\times10^{17}$~neutrons/cm$^2$ (60~h). The boron-10 isotope reacts with neutrons, undergoing transmutation to lithium-7 and alpha particles. Under these conditions, the resulting maximum lithium concentration in the h$^{10}$B$^{15}$N is $1.5\times10^{20}$~cm$^{-3}$ (60~h). The high energy nuclear reaction products also displaces boron and nitrogen atoms from their usual lattice site positions.

Both single-gated and dual-gated devices were constructed utilizing neutron-irradiated h$^{10}$B$^{15}$N dielectrics. The device fabrication procedure was the same as that used for the 5- and 24-layer devices. The single-gated device consisted of an irradiated h$^{10}$B$^{15}$N flake (60 hrs). The single-gated device had a graphite channel, and was stacked in the following order: normal hBN, graphite channel, irradiated h$^{10}$B$^{15}$N ($29$~nm), graphite gate. Two dual-gated bilayer graphene devices were constructed  utilizing irradiated h$^{10}$B$^{15}$N (10 hrs) as one of the gate dielectrics. The dual-gated devices were stacked in the following order, respectively. Device 1: graphite, normal hBN ($43$~nm), bilayer graphene, irradiated h$^{10}$B$^{15}$N ($33$~nm), graphite. Device 2: normal hBN, graphite, irradiated h$^{10}$B$^{15}$N ($42$~nm), bilayer graphene, normal hBN ($43$ nm), graphite. No intentional alignment between the graphite/bilayer graphene and hBN layers was performed during the stacking process.

Both the graphite and bilayer-graphene channel devices exhibit hysteresis upon sweeping the gate facing the irradiated h$^{10}$B$^{15}$N flake (Supplementary Information Fig.~\ref{ED_fig:tMM_carbon-defect-devices}c and Supplementary Information Fig.~\ref{ED_fig:tMM_IrBN-devices}). In contrast, no hysteresis is observed upon sweeping the gates facing the normal hBN in the bilayer graphene devices (Supplementary Information Fig.~\ref{ED_fig:tMM_IrBN-devices}a,d). Based on inspection of the graphite device and the dual-gate transport maps shown in Supplementary Information Fig.~\ref{ED_fig:tMM_IrBN-devices}c,f, we conclude that there is no GDW/ratchet effect. Instead, the role of the defects in the h$^{10}$B$^{15}$N dielectric appears to be to induce charge transfer with the channel and more conventional forms of hysteresis routinely seen in FETs with disordered dielectrics.

\textbf{Reproducibility of the GDW effect and thermal cycle dependence.} Although the observed GDW/ratchet phenomenology is reproducible upon sequentially sweeping gates (Supplementary Information Fig.~\ref{ED_fig:tMM_sweep_ranges}-\ref{ED_fig:reproducibility}), we find that there can be significant dependence on thermal cycling in our measurements. Supplementary Information Figs.~\ref{ED_fig:reproducibility}b-c show example measurements of $R_{xx}$ in the 24-layer device, taken across three different thermal cycles (i.e., completely removing the device from the cryostat and then reloading it again). In general, we see that the magnitude of the GDW hysteresis decreases upon subsequent thermal cycles (e.g. between Cycle 1 and 2). However, this is not always the case, as can be seen by comparing Cycles 2 and 3 where virtually no change in the magnitude of the GDW effect was observed. 

\section{Acknowledgements}
The authors thank Brian Skinner and Steven Simon for helpful discussions. This work was primarily supported by National Science Foundation (NSF) MRSEC 2308979. Device fabrication was partially supported by NSF CAREER award no. DMR-2041972. D.W. was supported by an appointment to the Intelligence Community Postdoctoral Research Fellowship Program at University of Washington administered by Oak Ridge Institute for Science and Education through an interagency agreement between the US Department of Energy and the Office of the Director of National Intelligence. M.Y. and X.X. acknowledge support from the State of Washington-funded Clean Energy Institute. Support for h$^{10}$B$^{15}$N crystal growth was provided by the Office of Naval Research, Award No. N00014-22-1-2582. The h$^{10}$B$^{15}$N neutron irradiation work was supported by the U.S. Department of Energy, Office of Nuclear Energy under DOE Idaho Operations Office Contract DE-AC07-05ID14517 as part of a Nuclear Science User Facilities award \#23-4683. K.W. and T.T. acknowledge support from the JSPS KAKENHI (Grant Numbers 21H05233 and 23H02052) and World Premier International Research Center Initiative (WPI), MEXT, Japan. This work made use of shared fabrication facilities at UW provided by NSF MRSEC 2308979.

\section{Author Contributions}
D.Waters, D.Waleffe, E.T. and E.A.-M. made the samples. D.Waters and D.Waleffe performed the measurements. J.F. assisted with optical spectroscopy characterization under the supervision of X.X. T.P. grew the irradiated h$^{10}$B$^{15}$N crystals under the supervision of J.E. K.W. and T.T. provided the remaining hBN crystals, including the hBN with large carbon defect concentration. M.Y. supervised the measurements. D.Waters, D.Waleffe, D.C. and M.Y. analyzed the data and wrote the manuscript.

\section*{Competing interests}
The authors declare no competing interests.

\section*{Additional Information}
Correspondence and requests for materials should be addressed to David Cobden or Matthew Yankowitz.

\section*{Data Availability}
Source data are available for this paper. All other data that support the findings of this study are available from the corresponding author upon request.

\bibliographystyle{naturemag}
\bibliography{references}

\newpage

\renewcommand{\figurename}{Supplementary Information Fig.}
\renewcommand{\thesubsection}{S\arabic{subsection}}
\setcounter{secnumdepth}{2}
\renewcommand{\thetable}{S\arabic{table}}
\setcounter{figure}{0} 
\setcounter{equation}{0}

\onecolumngrid
\newpage
\section*{Supplementary Information}

\begin{figure*}[h!]
\includegraphics[width=\textwidth]{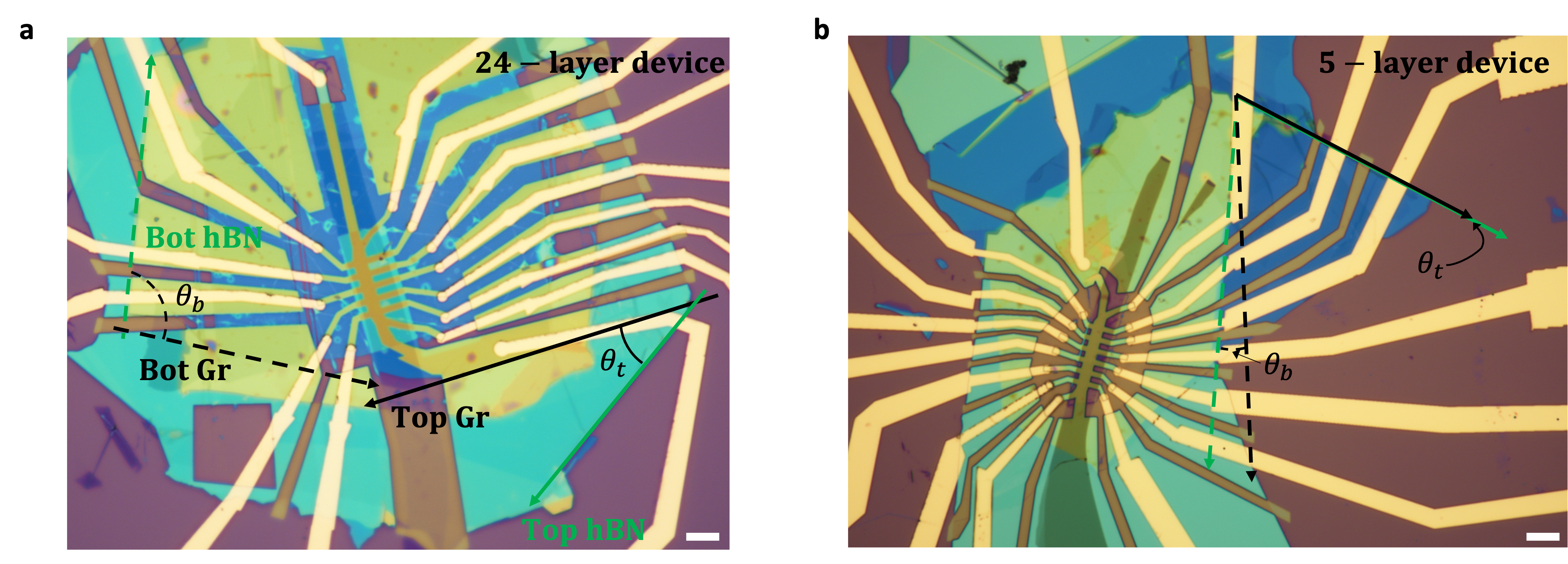} 
\caption{\textbf{Estimation of hBN alignment with the channel.} 
\textbf{a-b}, Optical micrographs of the 24-layer (\textbf{a})  and 5-layer (\textbf{b}) devices. Optically determined crystal axes are indicated for the top and bottom hBN by the green solid and dashed lines, respectively. The optically determined crystal axes are also shown for the top and bottom graphitic components of the twisted multilayer system as black solid and dashed lines, respectively. The angles between the top (bottom) hBN and top (bottom) graphite are denoted as $\theta_t$ ($\theta_b$). For the 24-layer device in \textbf{a}, $\theta_t = 2^{\circ} (\mathrm{mod}\ 30^{\circ})$ and  $\theta_b = 5^{\circ} (\mathrm{mod}\ 30^{\circ})$. For the 5-layer device in \textbf{b}, $\theta_t = 0.5^{\circ} (\mathrm{mod}\ 30^{\circ})$ and  $\theta_b = 6.5^{\circ} (\mathrm{mod}\ 30^{\circ})$. Scale bars are $5\ \rm{\mu m}$. 
}
\label{ED_fig:BN_alignment}
\end{figure*}

\begin{figure*}
\includegraphics[width=6in]{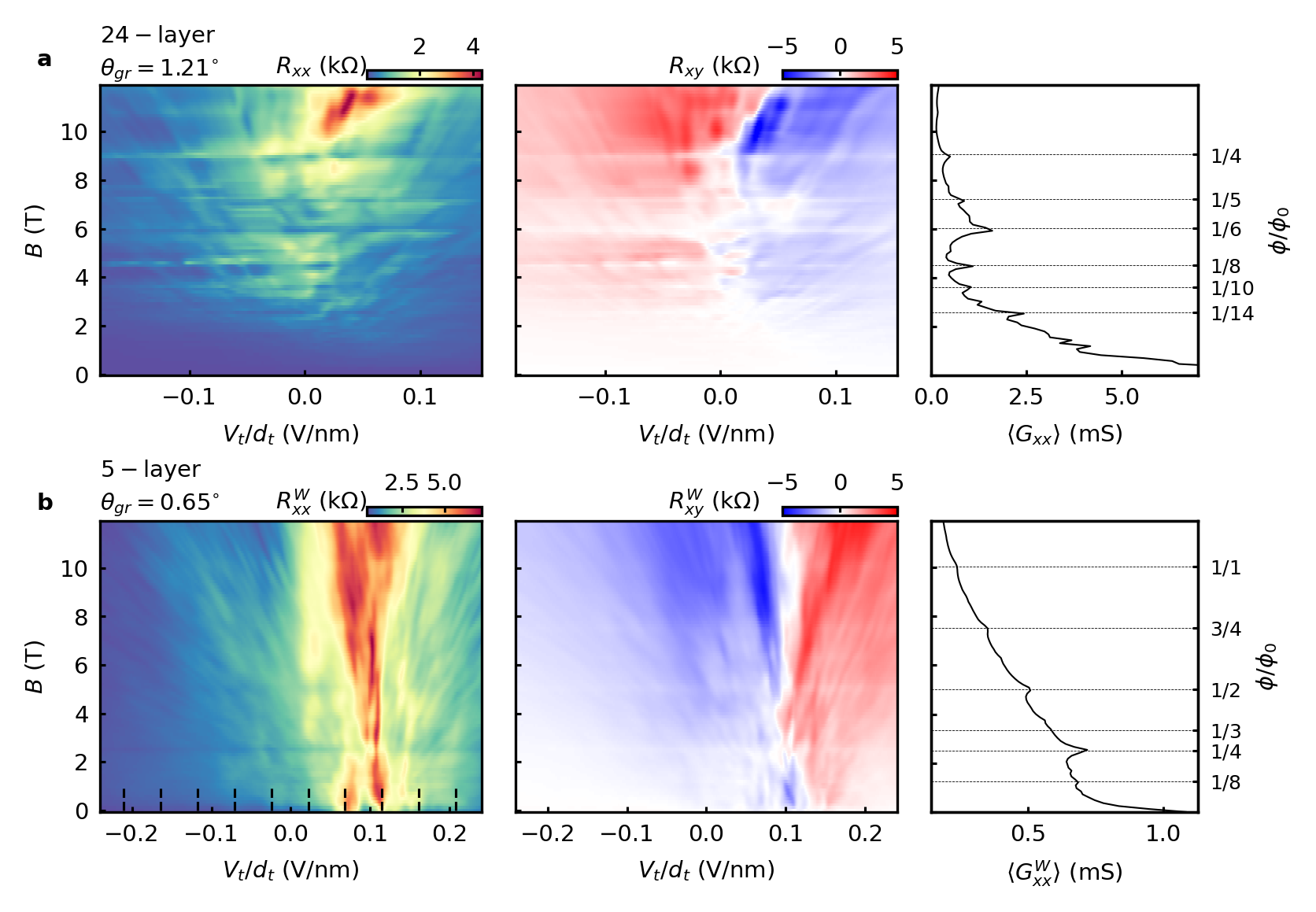} 
\caption{\textbf{Twist angle determination from Brown-Zak oscillations.}
\textbf{a}, Landau fan diagrams of $R_{xx}$ (left) and $R_{xy}$ (middle) from the 24-layer device as a function of the non-hysteretic gate voltage, $V_t$, and magnetic field, $B$, with $V_b=0$. Horizontal streaks in the data are Brown-Zak oscillations, arising due to the twisted interface between the two 12-layer graphene sheets. (Right) The conductance, $G_{xx}=R_{xx}/\left( R_{xx}^2 + R_{xy}^2\right)$, averaged over all values of $V_t$. The Brown-Zak oscillation peaks can be seen at rational fractions of the magnetic flux filling of the moir\'e lattice in units of the magnetic flux quantum, $\phi/\phi_0$, which allows us to extract the twist angle, $\theta_{gr}$ (Methods). 
\textbf{b}, Similar measurements for the 5-layer device. We show the measurements for the region of the device with the WSe$_2$ spacer, since these show the clearest superlattice features. In the leftmost panel, we also show vertical dashed lines that are spaced along the voltage axis by $\Delta n = C_t \Delta V_t/e = n_s$. The corresponding weak resistive bumps seen at low magnetic field are associated with fully filling several \moire mini-bands of the small-angle twisted interface. 
}
\label{ED_fig:twist_angle_tMM}
\end{figure*}

\begin{figure*}
\includegraphics[width=\textwidth]{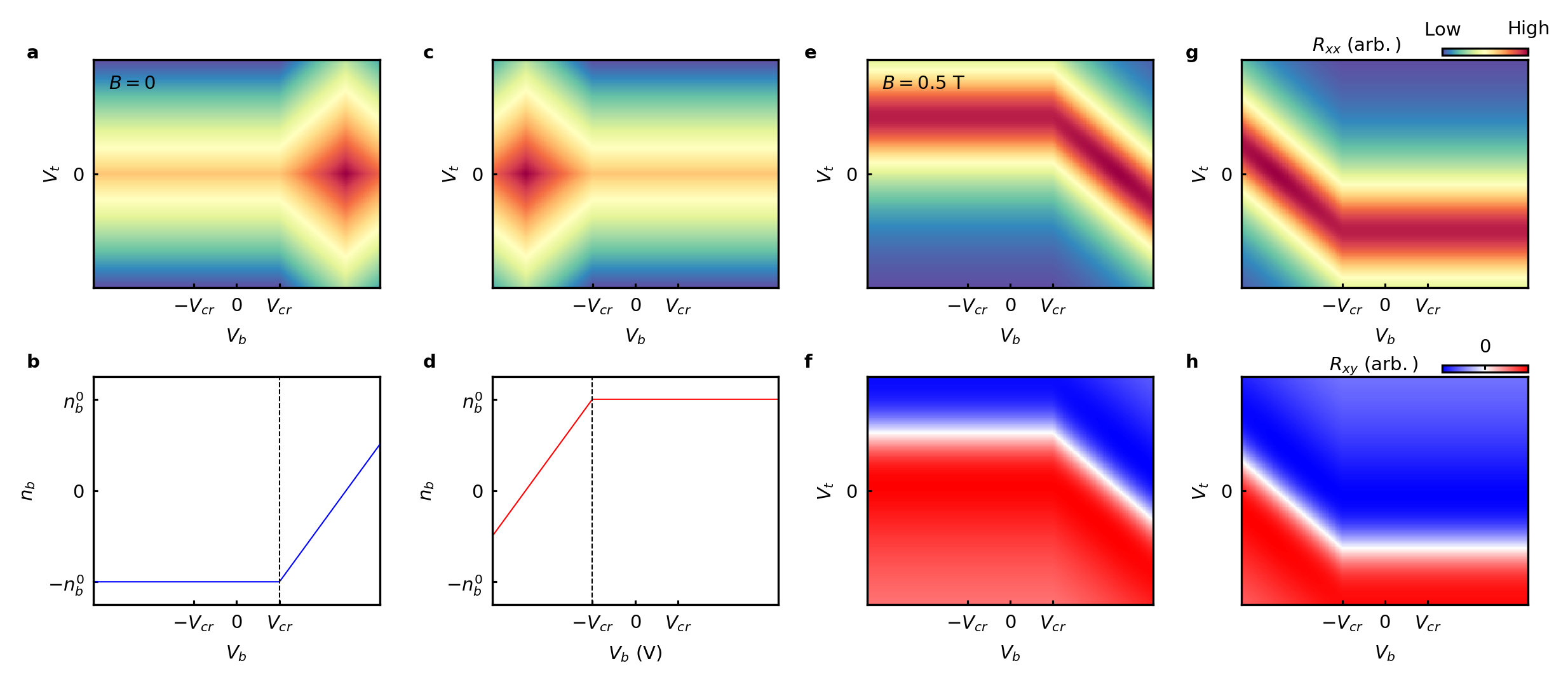} 
\caption{\textbf{Drude transport model for graphite including anomalous screening.}
\textbf{a}, Calculation of $R_{xx}$ at $B=0$ for a simple multi-component Drude model (see Methods). We artificially include the effect of anomalous screening of $V_b$ to capture the GDW effect. A cross-like pattern of the resistance features is seen in the map, similar to what is observed in Fig.~\ref{fig:1}. 
\textbf{b}, The density on the bottom surface, $n_b$, according to our phenomenological model. $n_b$ remains ``stuck'' at a hole density $-n_b^0$ until a critical voltage, $V_{cr}$ (dashed black line), after which the doping follows the ``normal'' field effect behavior. This pattern represents a typical forward sweep of the hysteretic gate in our 24-layer device. 
\textbf{c}-\textbf{d}, Similar plots for a typical backward sweep, where $n_b$ is ``stuck'' at an electron density $n_b^0$. 
\textbf{e-f}, Calculation of $R_{xx}$ (\textbf{e}) and $R_{xy}$ (\textbf{f}) from the same model but with $B=0.5~\rm T$, corresponding to the forward sweep (\textbf{b}). The cross-like feature seen at zero-field evolves into half of the parallelogram-like feature at $B=0.5$~T, reminiscent of the behavior seen in the 5-layer device (Fig.~\ref{fig:2}e-f). 
\textbf{g-h}, Similar plots for the backward sweep (\textbf{d)}.
}
\label{ED_fig:Drude_model}
\end{figure*}

\begin{figure*}
\includegraphics[width=4in]{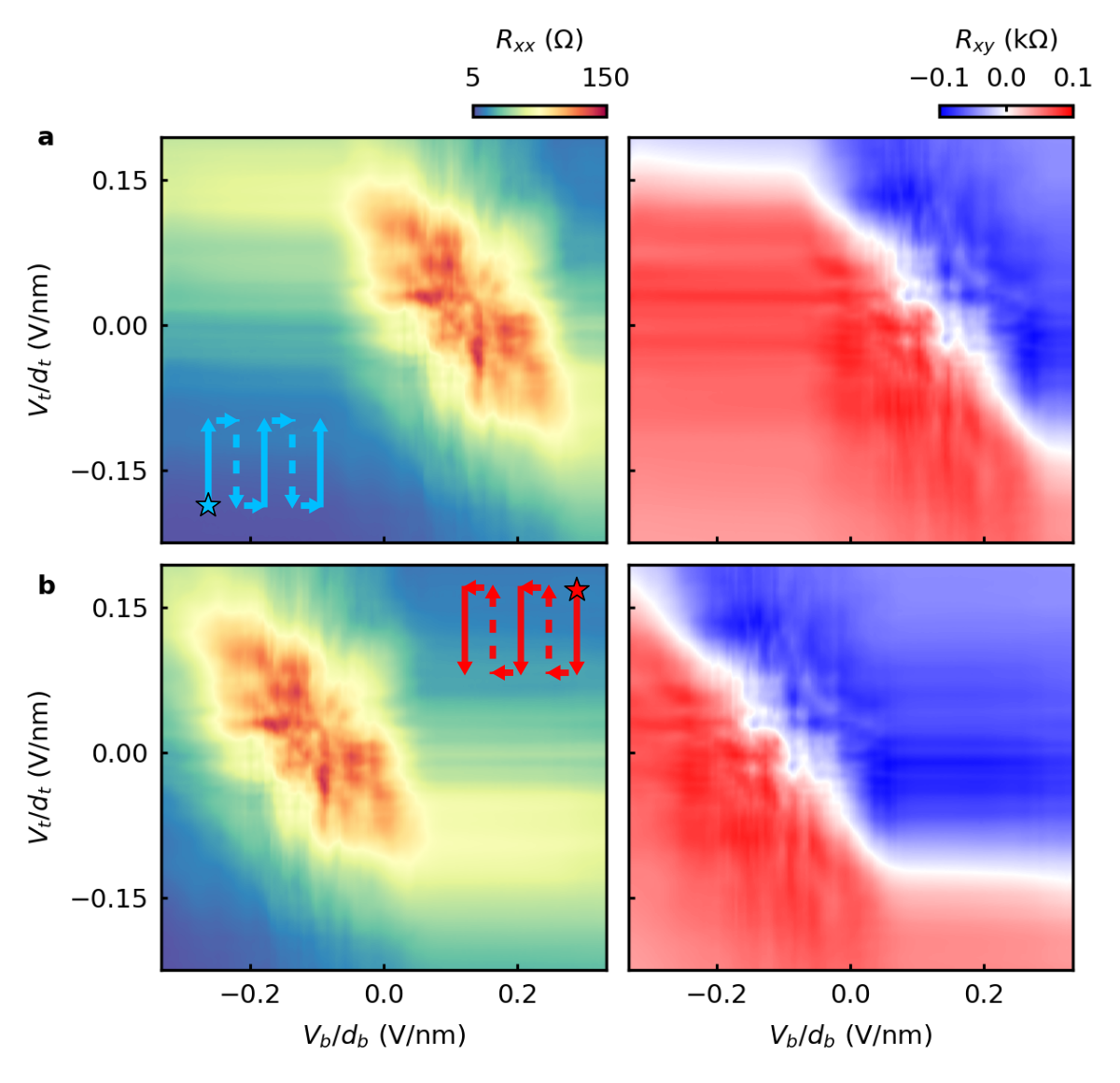} 
\caption{\textbf{Gate maps of the 24-layer device at $B=0.5$~T.} 
\textbf{a-b}, Dual-gate maps of $R_{xx}$ (left) and $R_{xy}$ acquired at $B=0.5$~T upon sweeping $V_t$ as the fast axis from negative to positive (\textbf{a}) and positive to negative (\textbf{b}). The observed behavior is captured by our simple Drude transport model (Methods and Supplementary Information Fig.~\ref{ED_fig:Drude_model}e-g). We see additional transport features arising from the twisted interface between the two 12-layer graphite flakes, which we do not attempt to model in Supplementary Information Fig.~\ref{ED_fig:Drude_model}.
}
\label{ED_fig:tMM_B_0p5T_maps}
\end{figure*}

\begin{figure*}
\includegraphics[width=\textwidth]{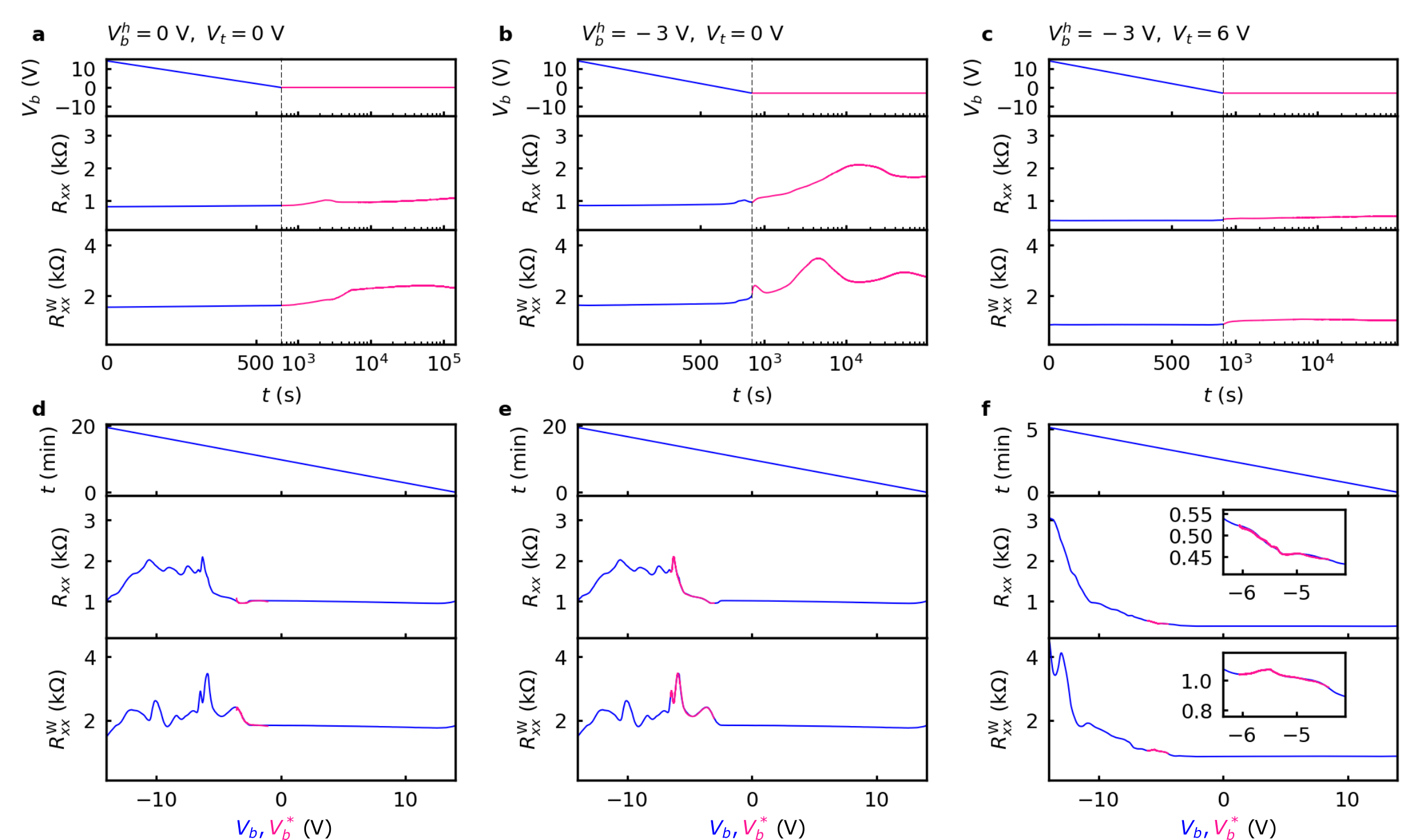} 
\caption{
\textbf{Timescale analysis with different measurement conditions}. 
\textbf{a-c}, Measurements of $R_{xx}$ and $R_{xx}^{\rm W}$ in the 5-layer device as a function of time (middle and bottom panels), under different initialization conditions. The top panel in each shows $V_b$ as a function of time. The vertical black dashed line indicates the time at which both gates are held at fixed values. In \textbf{a}, for example, $V_b$ is ramped from $-15~\rm V$ to a hold value of $V_b^h=0~\rm V$, while the top gate is held constant at $V_t=0~\rm V$. The analogous initialization conditions for \textbf{b-c} are listed above each panel.
\textbf{d-f}, Similar measurements as $V_b$ is swept from positive to negative with the same initalization conditions as \textbf{a-c}. The top panel shows the evolution of $V_b$ with time. The pink curves are the same data from the corresponding panels in \textbf{a-c}, converted from time to $V_b^*$ using the mapping described in the Methods. The same mapping is applied to the pink curves in the middle and bottom panels. The insets in the middle and bottom panels of \textbf{f} show a zoom-in on the relevant portion of the data, where it can be seen that the mapping clearly captures subtle features of the $R_{xx}$ and $R_{xx}^{\mathrm{W}}$ measurements. All of the mapping parameters are shown in Table~S1.
}
\label{ED_fig:timescales}
\end{figure*}

\clearpage
\begin{table}
\caption{Best fit parameters for the long-hold measurements.}
\begin{tabular}{| l | c | c | c | c |}
\hline
 Figure & $V_b^h~\rm (V)$ & $V_t~\rm (V)$ & $V_i~\rm (V)$ & $\tau_i~\rm (s)$\\ 
 \hline
 Fig. 3 & +3 & 0 & 2.14 & 6.60e2\\
 b, c & & & 0.71 & 1.29e4 \\
    & & & 1.43 & 3.30e5 \\
\hline
 SI Fig.~\ref{ED_fig:timescales} & 0 & 0 & -11.15 & 9.07e2 \\
  a, b  & & & -0.39 & 1.18e4 \\
   \hline
 SI Fig.~\ref{ED_fig:timescales}  & -3 & 0 & -2.08 & 1.34e1 \\
  c, d & & & -1.90 & 2.90e2 \\
   & & & -1.03 & 2.86e3 \\
   & & & -4.01 & 4.20e4 \\
   \hline
 SI Fig.~\ref{ED_fig:timescales}  & -3 & +6 & -2.40 & 3.72e2 \\
  e, f & & & -0.98 & 2.14e3 \\
   & & & -0.69 & 2.88e4 \\
   \hline
\end{tabular}
\end{table}
\clearpage

\begin{figure*}
\includegraphics[width=6.5in]{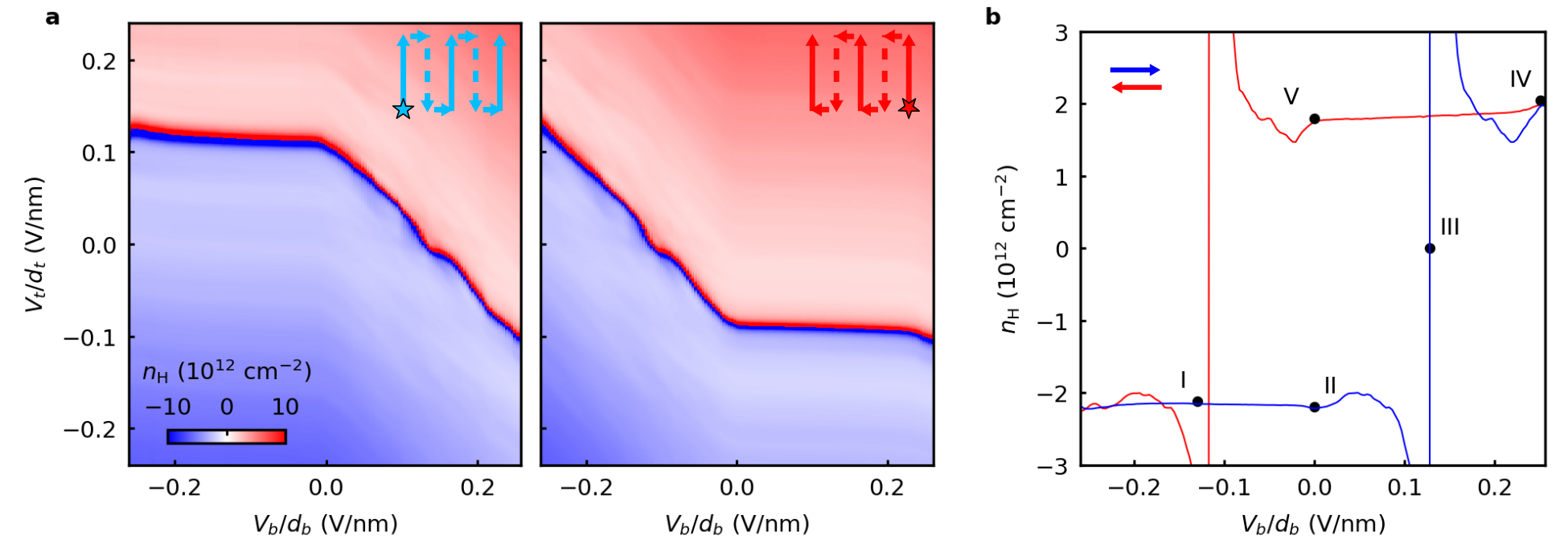} 
\caption{\textbf{Hall density in the 5-layer device.} 
\textbf{a}, Hall density, $n_H=B/eR_{xy}$, in the 5-layer device extracted from the Hall resistance data shown in Figs.~\ref{fig:2}e-f. The forward (left) and backward (right) sweep directions are indicated by the arrows in each panel. 
\textbf{b}, Line cuts taken at $V_t=0$ for both the forward (blue) and backward (red) maps. Points I-V correspond to the analogous markers in the schematic in Fig.~\ref{fig:4}. Points I and V correspond to the GDW regime, points II-IV corresponds to normal field effect. The Hall density diverges at the charge neutrality point (III) due to the complicated sequence of narrow \moire minibands arising due to the small value of $\theta_{gr}$ in the 5-layer device (i.e., bilayer graphene, with a simpler bandstructure, should instead form a diagonal line between points II and IV).
}
\label{ED_fig:Hall_density}
\end{figure*}

\begin{figure*}
\includegraphics[width=\textwidth]{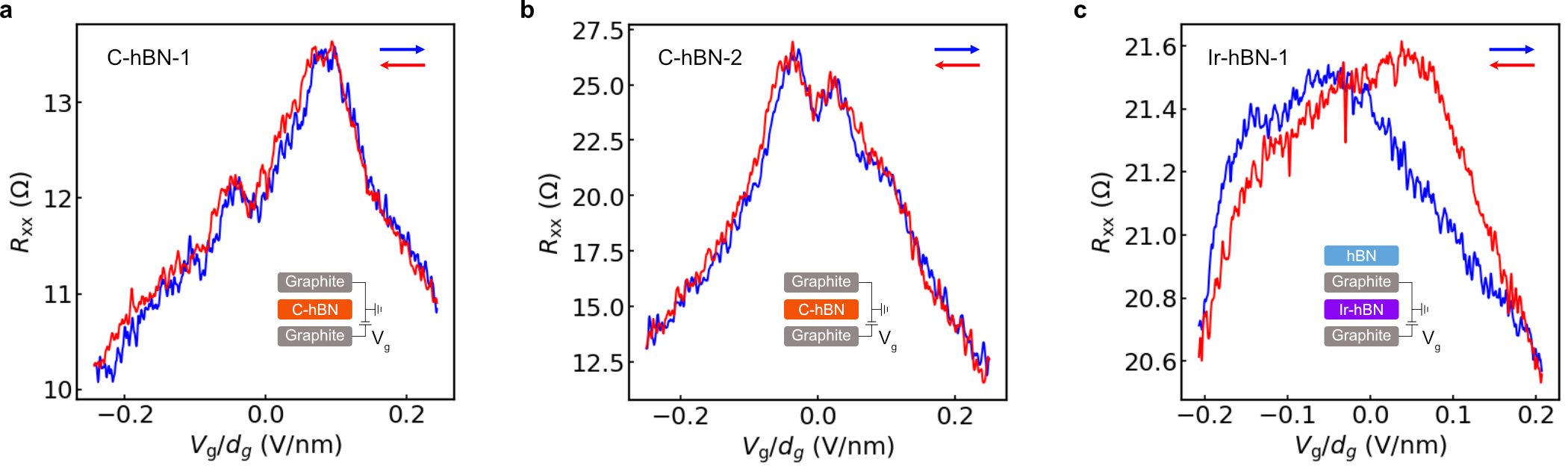} 
\caption{\textbf{Transport in graphite-channel devices with defect-rich hBN.} 
\textbf{a}, Measurement of $R_{xx}$ in a device consisting of two flakes of graphite encapsulating an hBN flake with a high concentration of carbon defects (C-hBN). Four-terminal transport is measured in the top graphene thin film, which is patterned into a Hall bar configuration.
\textbf{b}, Same measurement for a second device. No hysteresis is seen in either device. The measurements were taken at \textbf{a}, 350~mK and \textbf{b}, 4 K.
\textbf{c}, Same measurements for a device with a neutron-irradiated h$^{10}$B$^{15}$N dielectric (Ir-hBN). Neutron irradiation was performed for 60 hrs. Hysteresis is present, but there are no signatures of the GDW/ratchet effect. Measurements were taken at 4 K. The device has a capping hBN layer in order to facilitate easier transfer of the Ir-hBN.
}
\label{ED_fig:tMM_carbon-defect-devices}
\end{figure*}

\begin{figure*}
\includegraphics[width=\textwidth]{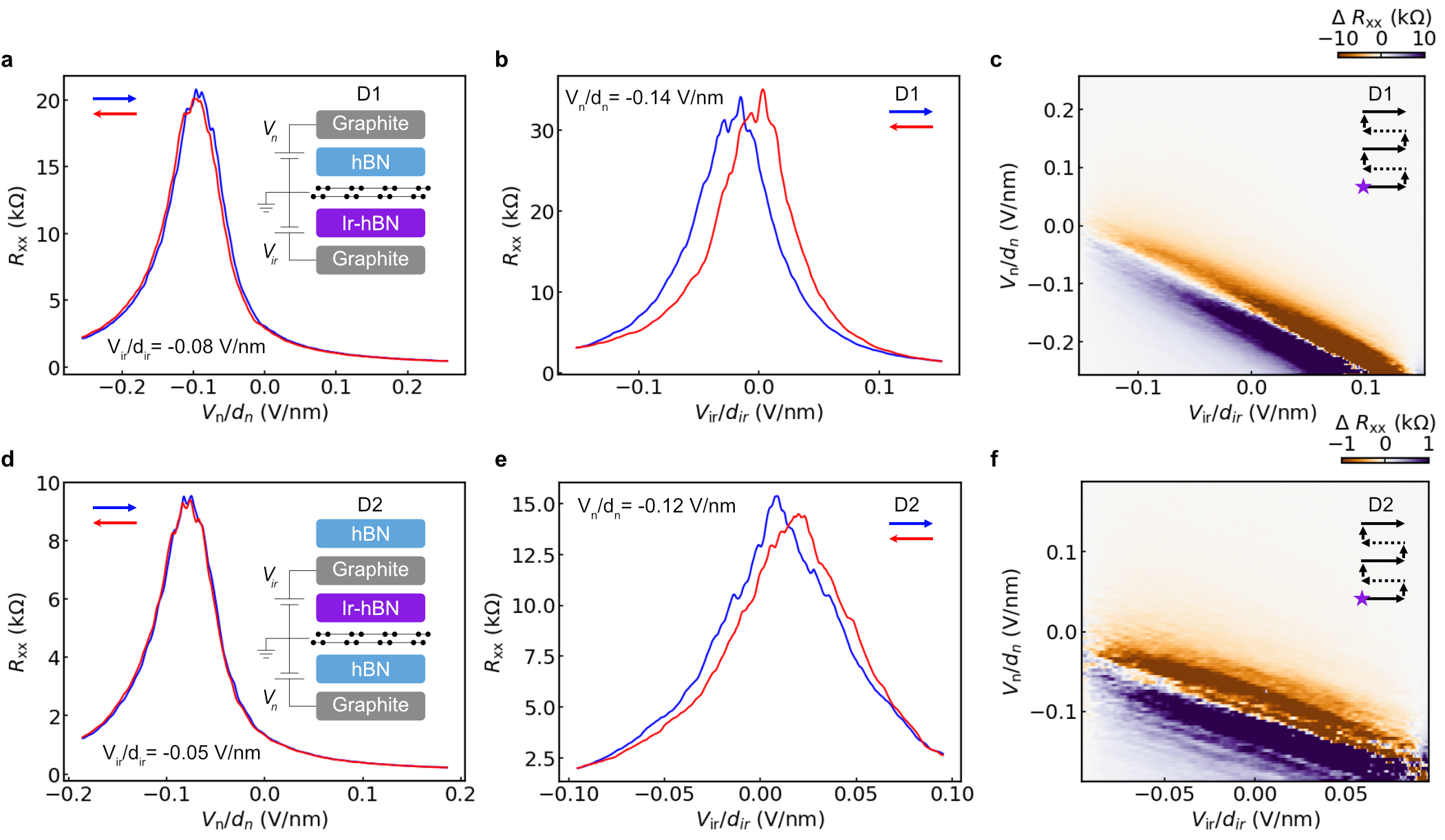} 
\caption{
\textbf{Transport in neutron-irradiated hBN devices with bilayer graphene channels}. 
\textbf{a-b}, Measurement of $R_{xx}$ in a dual-gated device with a Bernal bilayer graphene channel with a top dielectric made of ``normal'' hBN, and a bottom dielectric made of neutron-irradiated h$^{10}$B$^{15}$N, upon sweeping $V_n$ (\textbf{a}) and $V_{ir}$ (\textbf{b}). Neutron irradiation was performed for 10 hrs. The corresponding gates are denoted as $V_n$ and $V_{ir}$ respectively. There is substantial charge transfer induced by the Ir-hBN, as seen from the offset of the CNP to negative $V_n$ in \textbf{a}. There is no hysteresis upon sweeping $V_n$. There is hysteresis upon sweeping $V_{ir}$, but no signatures of the GDW/ratchet effect. 
\textbf{c}, Difference between the forward and backward scans, $\Delta R_{xx}=R_{xx}^{\rightarrow}-R_{xx}^{\leftarrow}$ for the same device when $V_{ir}$ is swept forward/backward at constant $V_n$. Arrows show the voltage path taken in order to create the map, and the star denotes the starting point of the measurement.  
\textbf{d-f}, Same as \textbf{a-c}, but for a second bilayer graphene device. D2 has a capping hBN layer which facilitated easier transfer of the Ir-hBN.
}
\label{ED_fig:tMM_IrBN-devices}
\end{figure*}

\begin{figure*}
\includegraphics[width=0.9\textwidth]{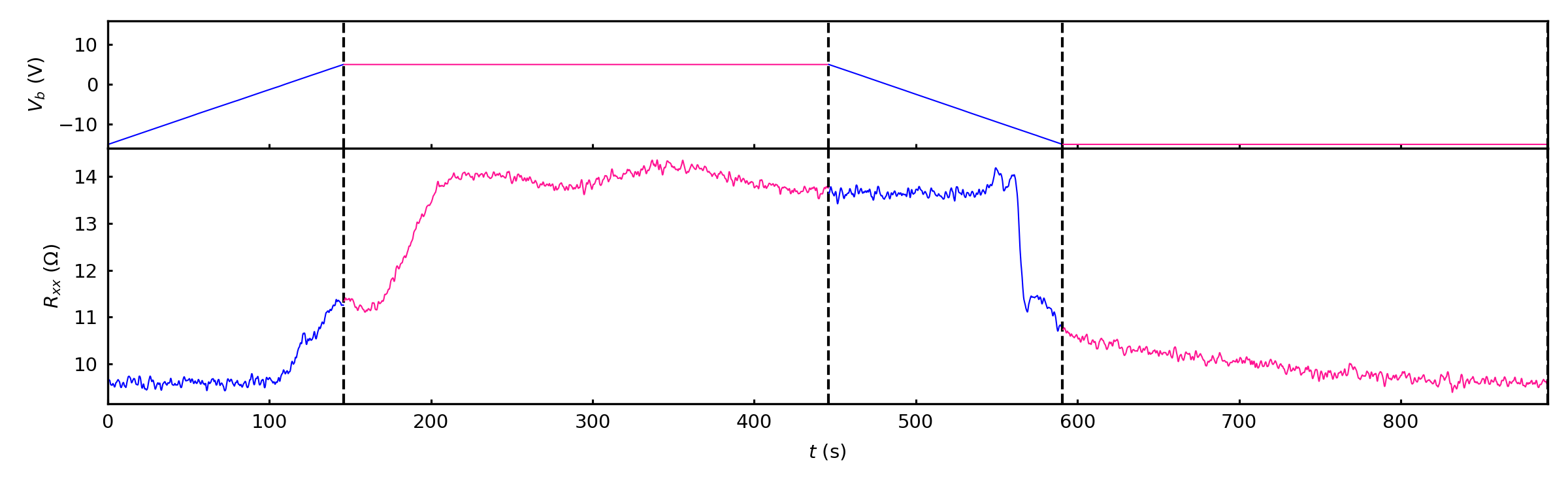} 
\caption{\textbf{Hold measurement in the 24-layer device.} 
(Bottom) $R_{xx}$ measured as a function of time in the 24-layer device. (Top) Voltage profile of the hysteretic gate, $V_b$, during the measurement. $V_t$ was held at ground. Blue curves indicate where $V_b$ is actively being ramped, whereas pink curves indicate when $V_b$ is held fixed. These regions are separated by the vertical dashed lines. $R_{xx}$ changes on minutes-long timescales even when $V_b$ is being held at a fixed value, consistent with our observations in the 5-layer device.
}
\label{ED_fig:tMM_hold_measurements}
\end{figure*}

\begin{figure*}
\includegraphics[width=6in]{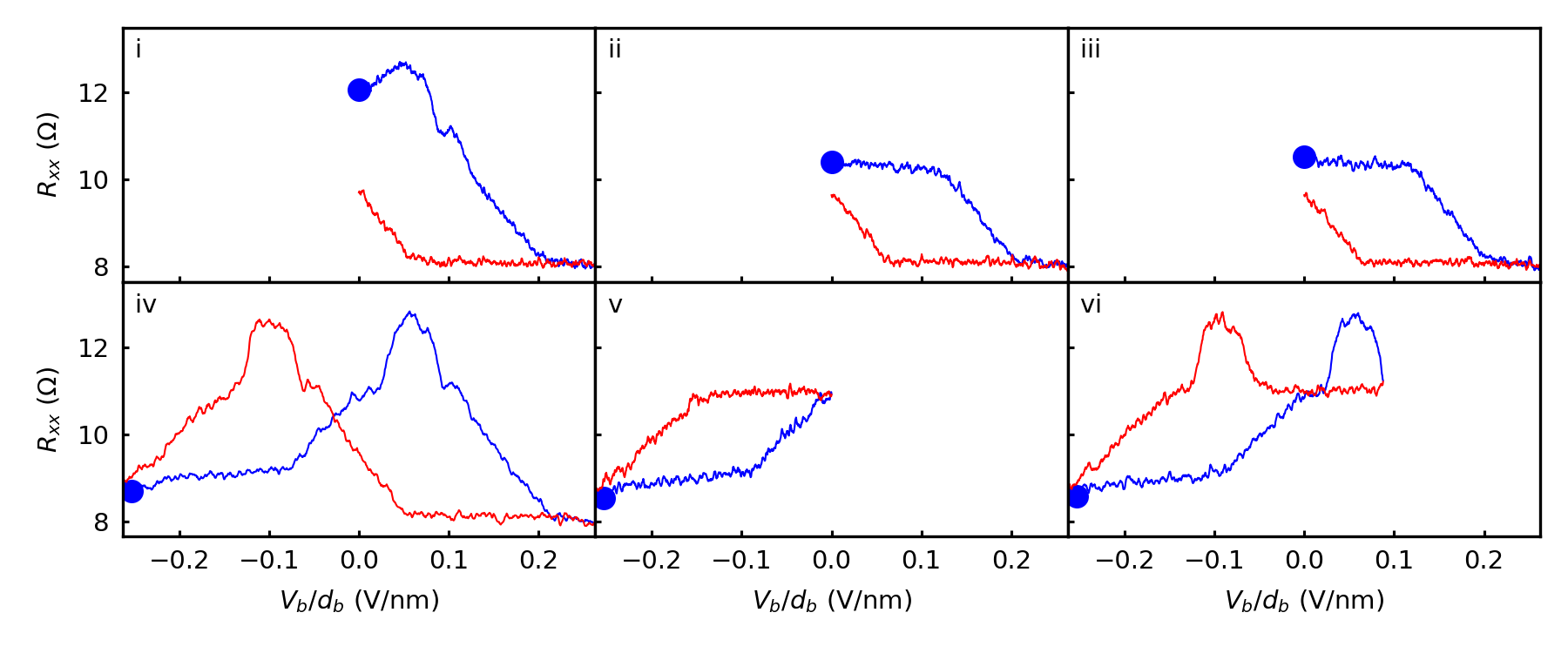} 
\caption{\textbf{Additional characteristics of the GDW/ratchet behavior.} $R_{xx}$ measured as the hysteretic gate, $V_b$, is swept in the 24-layer device. Each panels shows a different start, stop, and turning point in the gate sweep measurement. The starting point is denoted by the blue circle, the forward scan is shown in blue and the backward scan in red. Measurements were performed in sequential order, as denoted by the Roman numerals (i-vi) in the top left of each panel. The measurement in panel iii intentionally repeats the sweep range from panel ii, demonstrating the high degree of reproducibility of this behavior. There was a ramp step between measurements in panels iii and iv that is not shown. In each measurement, the GDW/ratchet effect is seen each time the sweep direction is reversed, regardless of the turning-point gate voltage. Data taken at $T=4~\rm K$. 
}
\label{ED_fig:tMM_sweep_ranges}
\end{figure*}

\begin{figure*}
\includegraphics[width=\textwidth]{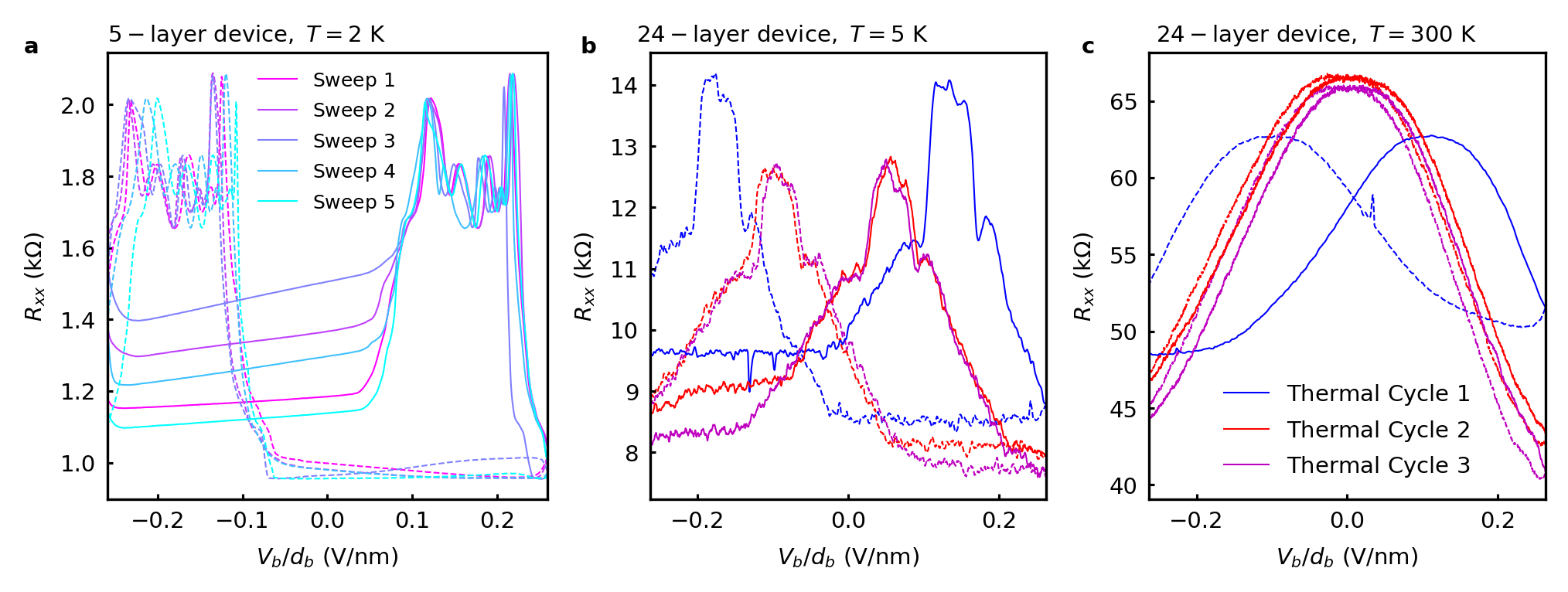} 
\caption{
\textbf{Reproducibility of the hysteresis characteristics}. 
\textbf{a}, $R_{xx}$ measured upon repeated sweeping of $V_b$ in the 5-layer device, showing that the qualitative behavior is reproducible. We note that there is slight variation in the magnitude of the hysteresis owing to the large-timescale charging behavior. 
\textbf{b-c}, $R_{xx}$ measured in the 24-layer device upon repeated thermal cycling. The measurements are performed at a temperature of 5~K in \textbf{b} and 300~K in \textbf{c}. For each thermal cycle, the measurement in \textbf{b} was taken first, then the sample was heated to $T=300~\rm K$ in order to take the measurement in \textbf{c}. For each panel in this figure, the solid (dashed) lines are the forward (backward) sweeping direction. For all sweeps, $V_t$, was held at ground. 
}
\label{ED_fig:reproducibility}
\end{figure*}

\begin{figure*}
\includegraphics[width=6.5in]{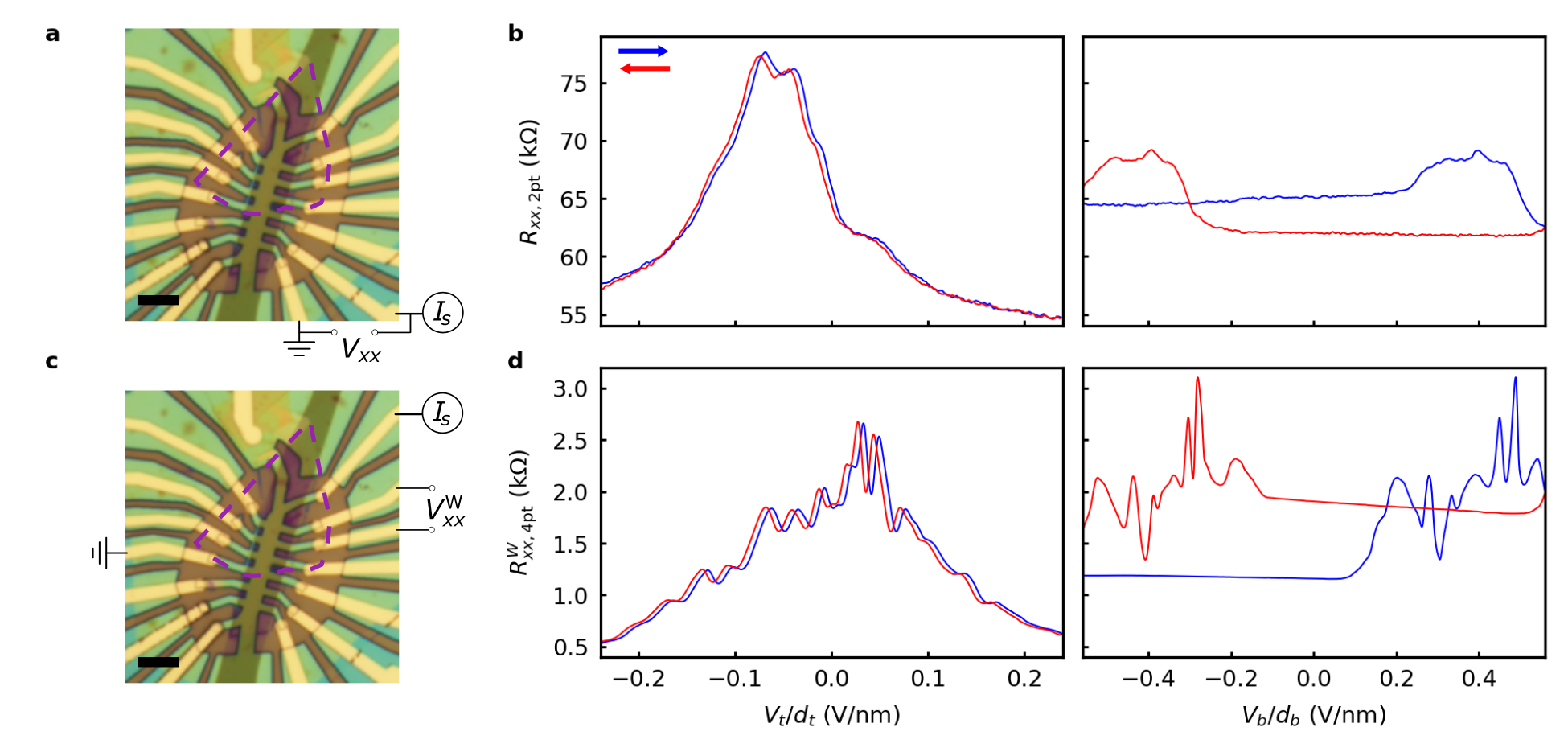} 
\caption{\textbf{Independent gate sweeps from the 5-layer device with and without the WSe$_2$ spacer.} 
\textbf{a}, Configuration of electrodes used to measure transport with current flowing only through the region of the device without the WSe$_2$ spacer. The limited number of contacts in this region of the device requires the use of a two-point (2pt) measurement configuration. We sourced current by applying a $1~\rm V$ a.c. voltage across a $10~\rm M\Omega$ resistor ($I_s=100~\rm nA$), then measured the potential difference across the source and drain pins, as shown. All other pins were floating to ensure that current only flows through the region of the device without the WSe$_2$ spacer. The resistance is then $R_{xx, \rm 2pt}=V_{xx}/I_s$, which includes contact resistance for this configuration.
\textbf{b}, Measured $R_{xx,2pt}$ upon sweeping $V_t$ (left) and $V_b$ (right), with sweep directions denoted by the arrows. Both are taken with the other gate grounded. The small offset seen in the $V_t$ sweep is due to the speed at which the gates are swept, and unrelated to the GDW/ratchet behavior seen upon sweeping $V_b$.
\textbf{c}, Similar to \textbf{a}, but showing a typical four-point resistance measure across only the region of the device with the WSe$_2$ spacer. All other pins are floated to ensure current only flows through the region of the device with the WSe$_2$ spacer. The resistance, \textbf{d}, is then $R_{xx,4pt}^{\rm W}=V_{xx}^{\rm W}/I_s$. 
\textbf{d}, Similar to \textbf{b}, but for the region of the device with the WSe$_2$ spacer. The results here show that the GDW/electron ratchet effects are seen even when current is passed only through a given portion of the device. Further, it shows that different contact pairs consistently show the GDW/electron ratchet effects, since the contact pairs shown here are different than those used for the rest of the figures.
}
\label{ED_fig:contact_comparison}
\end{figure*}

\end{document}